\documentclass[amsmath,amssymb,11pt,superscriptaddress,reprint, preprintnumbers, notitlepage,aps,twocolumn,nofootinbib]{revtex4-1}
\pdfoutput=1 
\usepackage[utf8]{inputenc}
\usepackage[english]{babel}
\usepackage{amsmath}
\usepackage{graphicx}
\usepackage{dcolumn}
\usepackage{pbox}
\usepackage{amssymb}
\usepackage{epsfig}
\usepackage{slashed}
\usepackage{amssymb}
\usepackage{ mathrsfs }
\usepackage{color}
\usepackage[font=small]{caption}
\usepackage[font=small]{subcaption}
\usepackage{url}
\usepackage{multirow}
\def\simlt{\stackrel{<}{{}_\sim}}

\usepackage{lineno}
\usepackage{xcolor,pifont}
\newcommand{\xmark}{\text{\ding{55}}}

\definecolor{MyLightBlue}{rgb}{0.22,0.51,0.9}
\definecolor{BrickRed}{rgb}{0.8, 0.25, 0.33}
\RequirePackage{hyperref}
\hypersetup{colorlinks, citecolor=blue,linkcolor=red, urlcolor=MyLightBlue}

\newcommand\scalemath[2]{\scalebox{#1}{\mbox{\ensuremath{\displaystyle #2}}}}

\makeatletter
\renewcommand\@makecaption[2]{%
  \par
  \vskip\abovecaptionskip
  \begingroup
  
   \small\rmfamily
    \begingroup
     \samepage
     \flushing
     \let\footnote\@footnotemark@gobble
     \@make@capt@title{#1}{#2}\par
    \endgroup
  \endgroup
  \vskip\belowcaptionskip
}
\makeatother

\DeclareUnicodeCharacter{2212}{-}
\begin{document}
\title{\Large  
Dark Matter and $(g-2)_{\mu,e}$ in radiative Dirac neutrino mass models
}

\author{\bf Talal  Ahmed  Chowdhury}
\email[E-mail: ]{talal@du.ac.bd}
\affiliation{Department of Physics, University of Dhaka, P.O. Box 1000, Dhaka, Bangladesh}
\affiliation{The Abdus Salam International Centre for Theoretical Physics, Strada Costiera 11, I-34014, Trieste, Italy}

\author{\bf Md. Ehsanuzzaman}
\email[E-mail: ]{ehsanzaman.k@gmail.com}
\affiliation{Department of Physics, University of Dhaka, P.O. Box 1000, Dhaka, Bangladesh}

\author{\bf Shaikh Saad}
\email[E-mail: ]{shaikh.saad@unibas.ch}
\affiliation{Department of Physics, University of Basel, Klingelbergstrasse\ 82, CH-4056 Basel, Switzerland}

\begin{abstract}
The origin of neutrino mass is a mystery, so is its nature, namely, whether neutrinos are Dirac or Majorana particles. On top of that, hints of large deviations of the muon and the electron anomalous magnetic moments (AMMs) are strong evidence for physics beyond the Standard Model. In this work, piecing these puzzles together, we propose a class of radiative Dirac neutrino mass models to reconcile $(g-2)_{\mu,e}$ anomalies with neutrino oscillation data. In this framework, a common set of new physics (NP) states run through the loops that generate non-zero neutrino mass and, due to chiral enhancement, provide substantial NP contributions to lepton AMMs. In addition, one of the three models studied in this work offers a Dark Matter candidate automatically stabilized by the residual symmetry, whose phenomenology is non-trivially connected to the other two puzzles mentioned above. Finally, our detailed numerical analysis reveals a successful resolution to these mysteries while being consistent with all colliders and cosmological constraints.
\end{abstract}

\maketitle
\section{Introduction}
The Standard Model (SM) of particle physics is the most successful theory in particle physics that describes the fundamental interactions between elementary particles. Despite its major triumph,  it is not perfect- it cannot explain  the origin of neutrino mass or dark matter (DM).  Moreover, the SM is under scrutiny since its predicted values of the muon and the electron anomalous magnetic moments\footnote{AMM is defined as $a_\ell=(g_\ell-2)/2$, where $\ell=e, \mu, \tau$.} (AMMs) are in tension with experimental measurements. There is a longstanding discrepancy  in the muon AMM measured  at BNL in 2006~\cite{Bennett:2006fi}. A new measurement performed at the Fermilab~\cite{Abi:2021gix} in 2021 is in excellent agreement with BNL's result, and combinedly they correspond to a large $4.2\sigma$ disagreement with the SM prediction~\cite{Aoyama:2020ynm} (for original works, see Refs.~\cite{Aoyama:2012wk,Aoyama:2019ryr,Czarnecki:2002nt,Gnendiger:2013pva,Davier:2017zfy,Keshavarzi:2018mgv,Colangelo:2018mtw,Hoferichter:2019mqg,Davier:2019can,Keshavarzi:2019abf,Kurz:2014wya,Melnikov:2003xd,Masjuan:2017tvw,Colangelo:2017fiz,Hoferichter:2018kwz,Gerardin:2019vio,Bijnens:2019ghy,Colangelo:2019uex,Blum:2019ugy,Colangelo:2014qya}):
\begin{align}
&\Delta a_\mu = (2.51\pm 0.59)\times 10^{-9}. \label{EXP-mu}
\end{align}

On the other hand, precise measurement of the fine-structure constant using Cesium atom at the Berkeley National Laboratory~\cite{Parker:2018vye} in 2018 yields,
\begin{align}
\alpha^{-1}(C_s) = 137.035999046(27).    
\end{align}
This result corresponds to a negative $2.4\sigma$ deviation of the electron AMM with respect to the SM value~\cite{Aoyama:2017uqe}:
\begin{align}
&\Delta a_e = (-8.8\pm 3.6)\times 10^{-13}. \label{EXP-e}
\end{align}
These discrepancies are large in magnitude, and the opposite sign between them is somewhat puzzling and hints towards physics beyond the SM (BSM). For attempts to solve these discrepancies simultaneously in BSM frameworks, see, e.g., Refs.~\cite{Giudice:2012ms, Davoudiasl:2018fbb,Crivellin:2018qmi,Liu:2018xkx,Dutta:2018fge, Han:2018znu, Crivellin:2019mvj,Endo:2019bcj, Abdullah:2019ofw, Bauer:2019gfk,Badziak:2019gaf,Hiller:2019mou,CarcamoHernandez:2019ydc,Cornella:2019uxs,Endo:2020mev,CarcamoHernandez:2020pxw,Haba:2020gkr, Bigaran:2020jil, Jana:2020pxx,Calibbi:2020emz,Chen:2020jvl,Yang:2020bmh,Hati:2020fzp,Dutta:2020scq,Botella:2020xzf,Chen:2020tfr, Dorsner:2020aaz, Arbelaez:2020rbq, Jana:2020joi,Chua:2020dya,Chun:2020uzw,Li:2020dbg,DelleRose:2020oaa,Kowalska:2020zve,Hernandez:2021tii,Bodas:2021fsy,Cao:2021lmj,Mondal:2021vou,CarcamoHernandez:2021iat,Han:2021gfu,Escribano:2021css,CarcamoHernandez:2021qhf,Chang:2021axw,Chowdhury:2021tnm,Bharadwaj:2021tgp,Borah:2021khc,Bigaran:2021kmn,PadmanabhanKovilakam:2022FN,Li:2021wzv,Biswas:2021dan,Julio:2022ton,Julio:2022bue}.

The observation of neutrino oscillations~\cite{Super-Kamiokande:1998kpq,Super-Kamiokande:2001ljr,SNO:2002tuh,KamLAND:2002uet,KamLAND:2004mhv,K2K:2002icj,MINOS:2006foh} was the first conclusive evidence that the SM is incomplete and must be extended. Although the existence of non-zero neutrino masses\footnote{For an extensive review on this subject, see  Ref.~\cite{Cai:2017jrq}.} has been firmly established, the nature of neutrinos, viz. Dirac or Majorana is still unknown. As widely known, observation of neutrinoless double beta decay (see, e.g., Ref.~\cite{Dolinski:2019nrj}) would settle this issue and establish the Majorana nature of neutrinos; however, all experiments so far have null results.  Similarly, despite the discovery that about eighty percent of the Universe's gravitating matter is non-luminous, we are yet to know anything about the nature of DM (see, e.g.,~\cite{Young:2016ala}).

\begin{figure}[b!]
\includegraphics[width=0.35\textwidth]{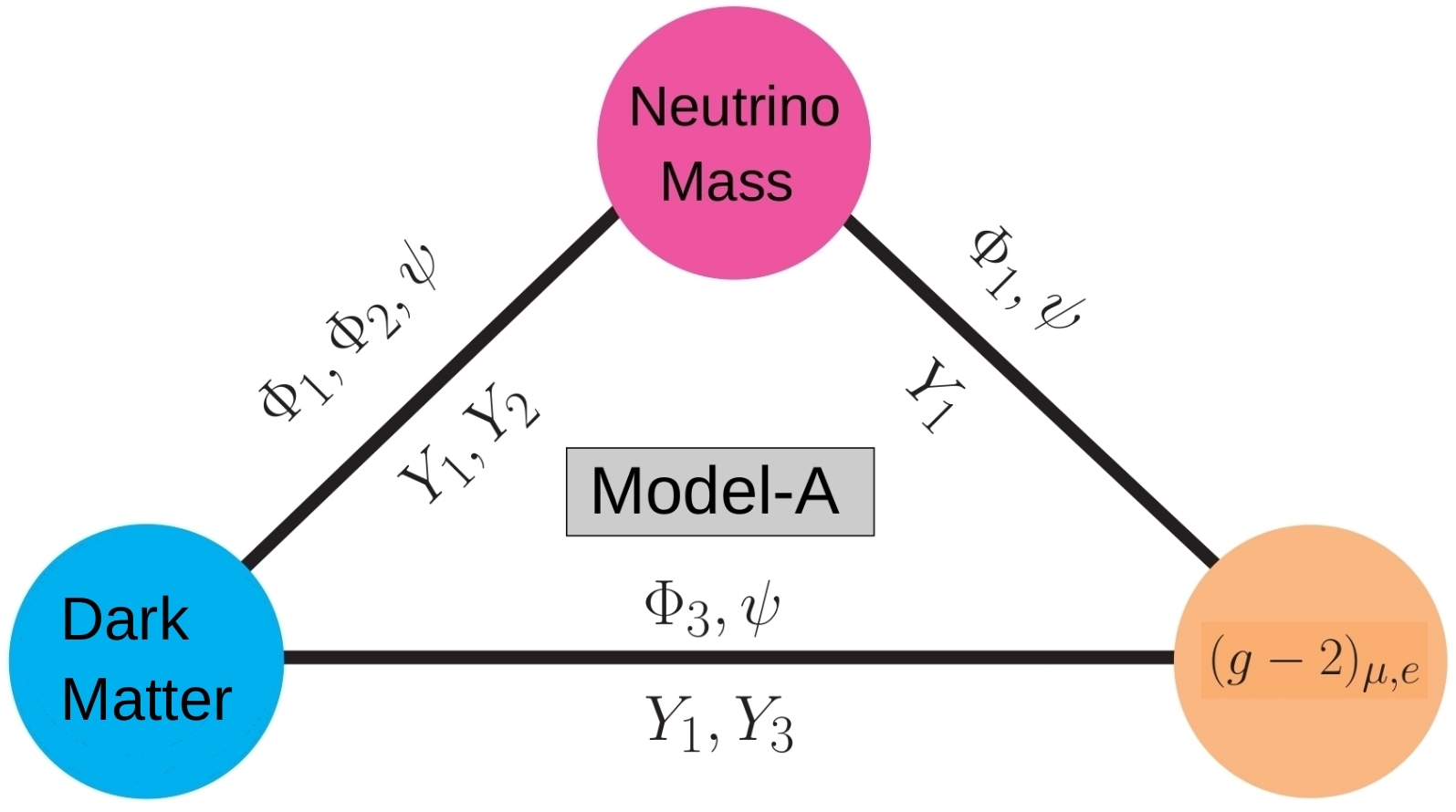}
\caption{Schematic diagram  demonstrating the link between neutrino mass generation, lepton $g-2$ anomalies, and Dark Matter in one of the benchmark models (Model-A) proposed in this work. See text for details.} \label{fig:00}
\end{figure}

This work considers neutrinos as Dirac particles, and non-zero neutrino masses originate from quantum corrections. This framework proposes a simultaneous solution to the muon and the electron AMMs that is non-trivially linked to the neutrino mass generation mechanism. New physics (NP) contributions to the lepton $g-2$ and non-zero neutrino masses arise via one-loop corrections mediated by a common set of BSM particles. We present a class of radiative Dirac neutrino mass models that share these same features and study in detail a particular model (dubbed as Model-A) belonging to this class that also addresses the DM puzzle and offers rich collider phenomenology. Remarkably, the stability of the DM is guaranteed by the residual symmetry that emerges after $U(1)_{B-L}$ gauge symmetry is spontaneously broken. In Model-A, all these puzzles are deeply intertwined, which is illustrated in Fig.~\ref{fig:00}.

The paper is organized as follows: in Sec.~\ref{sec:SETUP}, we discuss the general framework, and in Sec.~\ref{sec:MODEL}, we provide details of the set of models under investigation. Next, we discuss the experimental constraints in Sec.~\ref{sec:EXP} and carry out a detailed DM phenomenology for Model-A in Sec.~\ref{sec:DM}. Finally, we conclude in Sec.~\ref{sec:CONCLUSIONS}.

\section{Setup}\label{sec:SETUP}
Neutrinos being Dirac in nature requires the presence of right-handed partners $\nu_{R_i}$ ($i=1-3$), which automatically allows for tree-level neutrino mass via the term $-\mathcal{L}_y\supset Y_\nu \overline{L}\epsilon H^\ast \nu_R$ when the SM Higgs, $H$ acquires a vacuum expectation value (VEV);  here $L$ is the SM lepton doublet, and $\epsilon$ is the Levi-Civita tensor.  This, however, demands $y_\nu\sim \mathcal{O}(10^{-11})$ to be consistent with experimental data, which is seemingly unnatural~\cite{tHooft:1979rat} since the Yukawa couplings of the charged fermions in the SM are typically in the range $10^{-6}-1$. On the contrary, it is aesthetically attractive to generate Dirac neutrino mass radiatively that would naturally require the corresponding Yukawa couplings typically in the range $10^{-3}-1$. Symmetry arguments can naturally forbid the aforementioned tree-level term to achieve this. In this work, we accomplish this by extending the SM gauge symmetry by $U(1)_{B-L}$ \cite{Davidson:1978pm,Mohapatra:1980qe}; in the literature, various types of symmetries are imposed to realize radiative Dirac mass, see, e.g., Refs.~\cite{Mohapatra:1987hh, Mohapatra:1987nx, Balakrishna:1988bn, Branco:1978bz, Babu:1988yq, Gu:2007ug, Farzan:2012sa, Okada:2014vla, Ma:2016mwh, Bonilla:2016diq, Wang:2016lve, Ma:2017kgb, Yao:2017vtm, Wang:2017mcy, Helo:2018bgb, Reig:2018mdk, Han:2018zcn, Kang:2018lyy, Yao:2018ekp, Calle:2018ovc, CentellesChulia:2018gwr, Bonilla:2018ynb, Calle:2018ovc, Carvajal:2018ohk, CentellesChulia:2018bkz, Ma:2019yfo, Bolton:2019bou, Saad:2019bqf, Bonilla:2019hfb, Dasgupta:2019rmf, CentellesChulia:2019gic, CentellesChulia:2019xky, Jana:2019mez, Enomoto:2019mzl, Ma:2019byo,Restrepo:2019soi,Jana:2019mgj,Nanda:2019nqy,Wang:2020dbp,Borgohain:2020csn,Mahanta:2021plx,Bernal:2021ezl,Biswas:2021kio,Calle:2021tez,De:2021crr,Bernal:2021ppq,Mishra:2021ilq}.

Gauge anomaly cancellation conditions\footnote{As usual, quark fields $Q_L(3,2,1/6), u_R(3,1,2/3), d_R(3,1,-1/3)$ carry $1/3$ and SM leptons $L_L(1,2,-1/2), \ell_R(1,1,-1)$ carry $-1$ charges under $U(1)_{B-L}$ symmetry. The SM Higgs doublet $H(1,2,1/2)$ transforms trivially under $U(1)_{B-L}$.} then require the right-handed neutrinos to carry charges which are either $\nu_{R_{1,2,3}}=\{-1,-1,-1\}$ or $\nu_{R_{1,2,3}}=\{5,-4,-4\}$~\cite{Montero:2007cd, Machado:2010ui, Machado:2013oza}. We choose the latter charge assignment since the former allows the unwanted tree-level term in the Lagrangian. To spontaneously break  $U(1)_{B-L}$, we employ a SM singlet scalar that carries three units of $B-L$ charge: $\sigma\sim (1,1,0,3)$\footnote{Quantum number presented here is under the gauge group $\textrm{SM}\times U(1)_{B-L}=SU(3)_C\times SU(2)_L\times U(1)_Y\times U(1)_{B-L}$.}. Then non-zero  mass for the neutrinos appears through loop diagrams when both the electroweak (EW) and  $U(1)_{B-L}$ symmetries are broken (in our setup, the only two fields that acquire VEVs are $H$ and $\sigma$). These loop diagrams originate from ultra-violate (UV) completion of the following unique dimension five operator: 
\begin{align}
-\mathcal{L}_{d=5}= \frac{y_{ij}}{\Lambda}\;\overline{L}_i \epsilon H^\ast {\nu_R}_j\sigma+h.c., \label{d5}
\end{align}
where $i=1-3$ and $j=2,3$. In the following, we very briefly summarize how to construct UV-complete one-loop models utilizing this $d=5$ operator; for details, we refer the reader directly to  Ref.~\cite{Jana:2019mgj} (we adopt the nomenclature used therein). 

\begin{figure}[t!]
\includegraphics[width=0.47\textwidth]{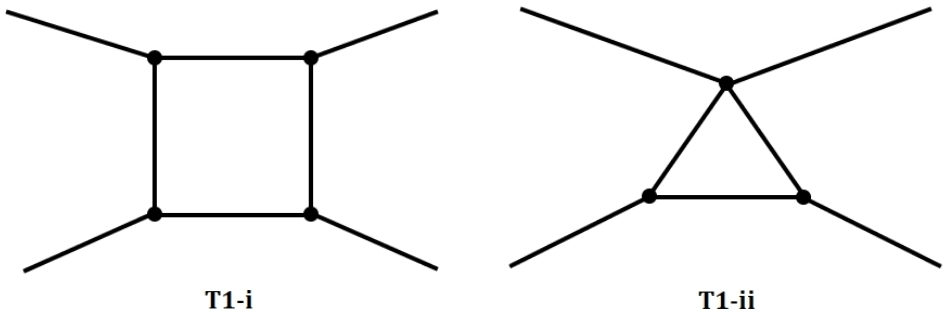}
\caption{Two distinct one-loop topologies that lead to Dirac neutrino mass. See text for details.} \label{fig:01}
\end{figure}
One-loop Dirac neutrino mass models can be constructed out of two independent \textit{topologies}: T1-i and T1-ii that are shown in Fig.~\ref{fig:01}. Depending on the  Lorentz structures (i.e, fermion-fermion-scalar or scalar-scalar-scalar interaction) associated with these vertices,  three different \textit{diagrams} (T1-i-1, T1-i-2, and T1-i-3) can be drawn for \textit{topology} T1-i. On the other hand, for T1-ii, there is a unique \textit{diagram} labeled as  T1-ii-1. Moreover, by interchanging the external scalar legs of some of these \textit{diagrams}, in total, eight minimal models can be fabricated~\cite{Jana:2019mgj}. 

In this work, we focus\footnote{Conclusions obtained in our work are very general and applicable to most of the  models fabricated from \textit{topology} T1-i.} on the \textit{diagram} T1-ii-1 and propose explicit models in light of the muon and the electron $g-2$. In particular, we  formulate three minimal models (labeled as Model-A, Model-B, Model-C), each of which, in addition to correctly reproducing neutrino oscillation data, addresses both $(g-2)_{\mu,e}$. Among these three models, DM can also be incorporated within Model-A. Moreover, this model shows profound correlations among the neutrino mass, DM, and $(g-2)_{\mu,e}$ as well as offers rich collider phenomenology.  Stunningly, no \textit{ad hoc} symmetry needs to be imposed by hand to realize this dark matter; instead, its stability is assured by a leftover discrete symmetry resulting from the breaking of $U(1)_{B-L}$ gauge symmetry.

\begin{figure}[th!]
\includegraphics[width=0.4\textwidth]{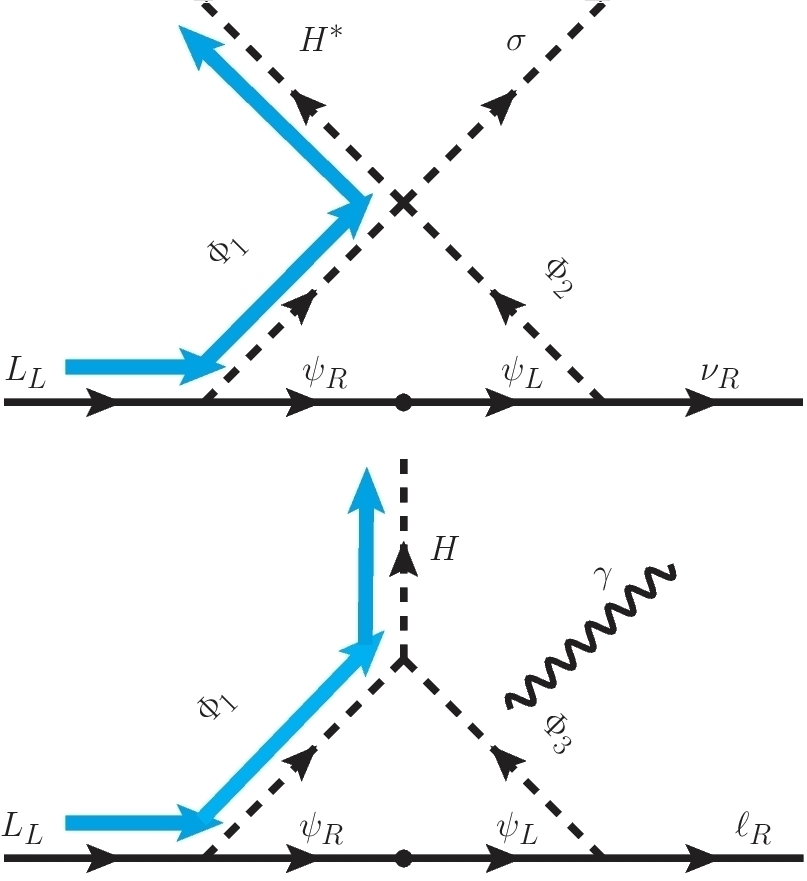}
\caption{ In this class of models,  neutrino mass (lepton $g-2$) originates from the diagram on the top (bottom). The blue line shows the direction of the iso-doublet flow in an economical fashion. Any other direction of this flow would correspond to non-minimal model, see text for details.} \label{fig:02}
\end{figure}

Since  \textit{topology} T1-i in Fig.~\ref{fig:01} has a 4-particle vertex, there is a unique choice of attaching external $H$ and $\sigma$ lines to it to form the $d=5$ operator  of Eq.~\eqref{d5}.   Completion of the neutrino mass diagram then requires (i) a vector-like\footnote{Due to vector-like nature, these fermions do not alter the anomaly cancellation conditions.} Dirac fermion $\psi$ (of three generations) and (ii) two distinct BSM scalars $\Phi_{1,2}$; see the top diagram in Fig.~\ref{fig:02}. With only these new states, corrections to muon and electron AMMs are too small to be consistent with experimental findings. However, large NP contributions to lepton $g-2$ can naturally arise within this setup via chirally enhanced terms that are proportional to vector-like fermion mass by introducing (iii) the third scalar $\Phi_3$; see the bottom diagram in Fig.~\ref{fig:02}. 

In Fig.~\ref{fig:02},  blue lines correspond to the iso-doublet flow  drawn to indicate the most economical way to build this class of models. As shown in Fig.~\ref{fig:02}, this choice requires a single BSM scalar field to be iso-doublet (and no BSM iso-doublet fermion is required). This way, the least number of new degrees of freedom is introduced in a given model  belonging to this class (this is our \textit{minimality} criterion).  Minimal models of this class then contain four BSM fields that propagate inside the loops and contribute to neutrino mass as well as lepton $g-2$, and  have the following quantum numbers:
\begin{align}
&\psi\sim (1,1,Y,\beta),\\
&\Phi_1\sim (1,2,-Y-\frac{1}{2},-\beta-1),\\
&\Phi_2\sim (1,1,Y,\beta+4),\\
&\Phi_3\sim (1,1,Y+1,\beta+1),
\end{align}
where, $Y$ and $\beta$ are the hypercharge and the $B-L$ charge, respectively, carried by the vector-like fermion. It is important to note that: (a) if $Y=-1$ then $\beta=2$ is not allowed. In this case, a cubic term of the form $\mathcal{V}\supset H^\dagger \Phi_1\sigma$ is allowed, which would lead to an induced VEV for $\Phi_1$ resulting in a tree-level Dirac mass for the neutrinos via $\mathcal{L}_y\supset  \overline{L}\epsilon \Phi_1^\ast \nu_R$. Furthermore, (b) if $Y=0$ then $\beta=-1$ is not allowed. In this scenario, a combination of three terms  $\mathcal{L}_y\supset \overline{L}\epsilon H^\ast \psi_R$, $\overline{\psi}_L\sigma\nu_R$, and $m_\psi \overline{\psi}_L \psi_R$  in the Lagrangian would generate neutrino mass via tree-level Dirac seesaw  (once VEVs of $H$ and $\sigma$ are inserted).

\section{Models}\label{sec:MODEL}
This section discusses three different versions of models belonging to the class introduced in the previous section. For simplicity, we restrict ourselves to the case of $|Y|\leq 1$. We label these models as Model-A ($Y=0$), Model-B ($Y=-1$), and Model-C ($Y=+1$) for which the full quantum numbers of NP states are specified in Table \ref{tab:particles}. In the following text, we provide all the necessary details of these models.

\begin{widetext}
\begin{table*}[th!]
\centering
{\footnotesize
\resizebox{0.9\textwidth}{!}{
\begin{tabular}{|c|c|c|c|}
\hline
{\bf Fields} & {\bf Model-A ($Y=0,\beta\neq -1$)}  & {\bf Model-B ($Y=-1,\beta\neq 2$)} & {\bf Model-C ($Y=+1$)}  \\[3pt]
\hline\hline
$\psi$&
$\psi^0(1,1,0,\beta)$ & $\psi^-(1,1,-1,\beta)$ & $\psi^+(1,1,+1,\beta)$ \\[3pt]
\hline
$\Phi_1$&
$\phi=\begin{pmatrix}\phi^0\\\phi^-\end{pmatrix}=(1,2,-\frac{1}{2},-\beta-1)$ & $\phi=\begin{pmatrix}\phi^+\\\phi^0\end{pmatrix}=(1,2,+\frac{1}{2},-\beta-1)$ & $\phi=\begin{pmatrix}\phi^-\\\phi^{--}\end{pmatrix}=(1,2,-\frac{3}{2},\beta-1)$       
 \\[3pt]
\hline
$\Phi_2$&
$S^0(1,1,0,\beta+4)$ & $\eta^-(1,1,-1,\beta+4)$ & $\eta^+(1,1,+1,\beta+4)$       
 \\[3pt]
\hline
$\Phi_3$&
$\eta^+(1,1,+1,\beta+1)$ & $S^0(1,1,0,\beta+1)$  & $\kappa^{++}(1,1,+2,\beta+1)$      
 \\[3pt]
\hline \hline
DM?&
$\psi^0$ & $\xmark$  & $\xmark$      
 \\[3pt]
\hline
\end{tabular}
}
    \caption{ Particle contents of new physics models proposed in this work. }
    \label{tab:particles}
    }
\end{table*}
\end{widetext}

\textbf{Yukawa interactions:}-- In three of these models, the new Yukawa part of the Lagrangian takes the following general form:
\begin{align}
-\mathcal{L}_Y&= Y^{iJ}_1 \overline L_{L_i} \Phi_1 \psi_{R_J} +Y^{Ij}_2 \overline \psi_{L_I} \Phi_2 \nu_{R_{ j\neq3}}   
\nonumber\\
&+Y^{Ij}_3 \overline \psi_{L_I}\Phi_3\ell_{R_j}+ M^{IJ}_F\overline \psi_{L_I}\psi_{R_J}.
\end{align}
Here $Y_{1,2,3}$ are in general $3\times 3$ arbitrary matrices, and we define their entries by,
\begin{align}
&Y_1^{ij}=y_{ij},\;Y_2^{ij}=x_{ij},\;Y_3^{ij}=z_{ij}.    
\end{align}
Without loss of generality, we choose to work in a basis where the vector-like fermion mass matrix is diagonal, 
\begin{align}
M_F=diag(M_1,M_2,M_3).    
\end{align}
From Fig.~\ref{fig:02}, it can be seen that $Y_1$ and $Y_2$ are responsible for neutrino mass generation, whereas, $Y_1$ and $Y_3$ provide NP contributions to the lepton $g-2$ that are chirally  enhanced. For sizable Yukawa couplings, lepton flavor violating  (LFV) processes provide stringent constraints on the off-diagonal couplings of these matrices and force them to take almost diagonal form. To be consistent with the experimental data of $(g-2)_{e,\mu}$, entries of $Y_1$ and $Y_3$ coupling matrices are required to be substantial; hence to suppress LVF,  we adopt diagonal textures for these two matrices. On the other hand, entries of $Y_2$ are required to be somewhat smaller to incorporate correct neutrino mass scale.   Hence, for the rest of the analysis, the Yukawa coupling matrices are chosen to  have the following form:
\begin{align}
&Y_1=\scalemath{0.9}{
\begin{pmatrix}
y_1&0&0\\
0&y_2&0\\
0&0&y_3
\end{pmatrix} },
Y_2=\scalemath{0.9}{
\begin{pmatrix}
0&x_{12}&x_{13}\\
0&x_{22}&x_{23}\\
0&x_{32}&x_{33}
\end{pmatrix} },
Y_3=\scalemath{0.9}{
\begin{pmatrix}
z_1&0&0\\
0&z_2&0\\
0&0&z_3
\end{pmatrix} }.\label{Ys}
\end{align}
Note that, due to different $B-L$ charge assignments of the right-handed neutrinos, the first column of $Y_2$ is zero. 
For the simplicity of our work, we treat all couplings to be real.

\textbf{Scalar interactions:}--
Owing to the $B-L$ charge assignments, the  scalar potential of this theory takes a  simple form. Instead of writing the entire potential, we only provide the relevant interactions required to generate neutrino mass as well as lepton AMMs,  
\begin{align}
-V_{\nu}&\supset
\lambda H\epsilon\Phi_1\Phi_2 \sigma^\ast + \mu H^\dagger  \Phi_1 \Phi_3  + h.c.\;, \label{mix}
\end{align}
here, $\epsilon$ is the Levi-Civita tensor. Since the SM Higgs doublet transforms trivially under $B-L$, it does not mix with the BSM scalars. Its VEV $\langle H\rangle=v_H/\sqrt{2}$ (where $v_H=246$ GeV) breaks the electroweak (EW) symmetry, while VEV of $\sigma$ field $\langle \sigma\rangle=v_\sigma/\sqrt{2}$ breaks the $U(1)_{B-L}$  symmetry. As a result of these spontaneous symmetry breakings, BSM neutral (charged) scalars originating from $\Phi_{1,2,3}$ mix as can be seen from Eq.~\eqref{mix}. Then the corresponding mass-squared matrices can be written as,   
\begin{align}
V&\supset 
\begin{pmatrix}
\phi^{0 \ast}&S^0
\end{pmatrix}
\begin{pmatrix}
m^2_\phi&a_0\\
a_0&m^2_S
\end{pmatrix}
\begin{pmatrix}
\phi^{0}\\S^{0\ast}
\end{pmatrix}
\nonumber \\&+
\begin{pmatrix}
\phi^+&\eta^+
\end{pmatrix}
\begin{pmatrix}
m^2_\phi&a_+\\
a_+&m^2_\eta
\end{pmatrix}
\begin{pmatrix}
\phi^-\\ \eta^-
\end{pmatrix}
\nonumber \\&+
\begin{pmatrix}
\phi^{++}&\kappa^{++}
\end{pmatrix}
\begin{pmatrix}
m^2_\phi&a_{++}\\
a_{++}&m^2_\kappa
\end{pmatrix}
\begin{pmatrix}
\phi^{--}\\ \kappa^{--}
\end{pmatrix},
\end{align}
where we have defined the following quantities: 
\begin{align}
\textrm{Model-A:}&\; \textrm{(no doubly charged scalar)}\nonumber \\& 
a_0=-\frac{\lambda}{2}v_Hv_\sigma,\;
a_+=\frac{\mu}{\sqrt{2}}v_H,\;
\\
\textrm{Model-B:}&\; \textrm{(no doubly charged scalar)}\nonumber \\& 
a_0=\frac{\mu}{\sqrt{2}}v_H,\;
a_+=-\frac{\lambda}{2}v_Hv_\sigma,\;
\\
\textrm{Model-C:}&\; \textrm{(no neutral scalar)}\nonumber \\& 
a_+=-\frac{\lambda}{2}v_Hv_\sigma,\;
a_{++}=\frac{\mu}{\sqrt{2}}v_H.
\end{align}
Furthermore, we diagonalize these matrices as,
\begin{align}
&M^2_x=O_x\; \textrm{diag}\{(M^x_1)^2, (M^x_2)^2\}\;O^T_x,
\\
&O_x=\begin{pmatrix}
\cos\theta_x&-\sin\theta_x\\
\sin\theta_x&\cos\theta_x
\end{pmatrix},
\\
&\sin 2\theta_x=\frac{2a_x}{(M^x_1)^2-(M^x_2)^2},
\label{scalarmassmix}
\end{align}
where we use the notation: $x=\{0,+,++\}$ and $M_1>M_2$. We denote these two mass eigenstates  $S^x_{1,2}$ by  (i) $S^0_i\equiv S_i$ for neutral, (ii) $S^+_i\equiv \eta_i$ for singly-charged, and (iii) $S^{++}_i\equiv \kappa_i$ for doubly-charged scalars. Explicitly, the flavor and the mass eigenstates are related via the following identities:  
\begin{align}
\textrm{neutral:}\;\; &\phi^{0\ast}=c_{\theta_0}S_1 - s_{\theta_0}S_2,\\
&S^{0}=s_{\theta_0}S_1 + c_{\theta_0}S_2,\\
\textrm{singly-charged:}\;\; &\phi^{+}=c_{\theta_+}\eta_1 - s_{\theta_+}\eta_2,\\
&\eta^{+}=s_{\theta_+}\eta_1 + c_{\theta_+}\eta_2,\\
\textrm{doubly-charged:}\;\; &\phi^{++}=c_{\theta_{++}}\kappa_1 - s_{\theta_{++}}\kappa_2,\\
&\kappa^{++}=s_{\theta_{++}}\kappa_1 + c_{\theta_{++}}\kappa_2.
\end{align}

It is important to note that due to the simplified form of the scalar potential, there is  no mass splitting between $Re[\phi^0]$ and $Im[\phi^0]$ components; this is why, neutral scalar cannot serve as a viable DM candidate in Model-A and -B (Model-C does not contain any neutral scalar within $\Phi_{1,2,3}$). Consequently, the only model that provides a DM candidate is Model-A (see Sec.~\ref{sec:DM} for details), which is a Dirac fermion DM (Model-B and Model-C do not contain electrically neutral BSM fermion).

\textbf{Neutrino mass:}-- 
The leading contributions to neutrino masses in this theory appear at the one-loop order, as shown in Fig.~\ref{fig:02} (Feynman diagram on the top). In this Feynman diagram, BSM neutral (singly-charged) scalars and fermions run through the loop in Model-A (Model-B and Model-C). It is straightforward to compute the neutrino mass formula, which is given by,   
\begin{align}
M^{ij}_\nu&=
\frac{\sin 2\theta_x}{8\pi^2} \left(Y^{ik}_1\right)^\ast M_{F_k} \left(Y^{kj}_2\right)^\ast
\nonumber \\&\times 
\bigg\{ \frac{M^2_{S^x_2}\log\left[ \frac{M_{F_k}^2}{M^2_{S^x_2}} \right]}{M^2_{F_k}-M^2_{S^x_2}} - \left(S^x_2\to S^x_1 \right) \bigg\} 
\nonumber \\ &
\equiv Y_1^\ast \hat M Y_2^\ast,
\end{align}
where,
\begin{align}
&\hat M= \hat M_i \delta_{ij}, 
\\&
\hat M_k= \frac{\sin 2\theta_x}{8\pi^2}  M_{F_k}
\bigg\{ \frac{M^2_{S^x_2}\log\left[ \frac{M_{F_k}^2}{M^2_{S^x_2}} \right]}{M^2_{F_k}-M^2_{S^x_2}} - \left(S^x_2\to S^x_1 \right) \bigg\}, 
\end{align}
and
$x=0, +, +$ for Model-A, -B, -C, respectively. Since neutrinos are Dirac particles,  it is simple to solve for the Yukawa couplings $x_{ij}\in Y_2$ in  terms of neutrino observables that are known quantities and in terms of couplings $y_i$ (to be determined from lepton $g-2$) as follows:   
\begin{align}
&x_{ij}=\frac{U_{ij}m_j}{y_i \hat M_i},   \label{x-sol} 
\end{align}
here
\begin{align}
&m_1=0,\;\; m_2=\sqrt{ \Delta m^2_\textrm{sol} }, \;\;
m_3=\sqrt{  \Delta m^2_\textrm{atm} },
\end{align}
and $U$ is the left-rotation matrix that diagonalizes the neutrino mass matrix (recall that Dirac neutrino mass matrix is not symmetric), i.e.,
\begin{align}
U^\dagger m_\nu m_\nu^\dagger U=\left( m^\textrm{diag}_\nu \right)^2.    
\end{align}
The solution given in Eq.~\eqref{x-sol} corresponds to normal mass ordering for neutrinos. 
Analogously, the solution for inverted ordering can be trivially constructed by relabelling the $B-L$ charges of the right-handed neutrinos, which would correspond to the third column being zero in Eq.~\eqref{Ys} instead of the first column.

Note that, due to the non-universal  charge assignments of the right-handed neutrinos, one of them carrying five units of $B-L$ charge remains massless (as well as the lightest SM neutrino). However, within this framework, non-zero $m_1$ is generated via dimension-7 operator\footnote{An UV-completion of this dimension-7 operator requires two more copies of  $\Phi_2$-like fields: $\Phi_2^{\prime}(1,1,Y,\beta-5)$ and $\Phi_2^{\prime\prime}(1,1,Y,\beta-2)$.} of the form $\mathcal{L}_7\supset \overline L_L H^\ast \nu_R \sigma^\ast \sigma^\ast/\Lambda^3$.

\begin{figure}[th!]
\includegraphics[width=0.47\textwidth]{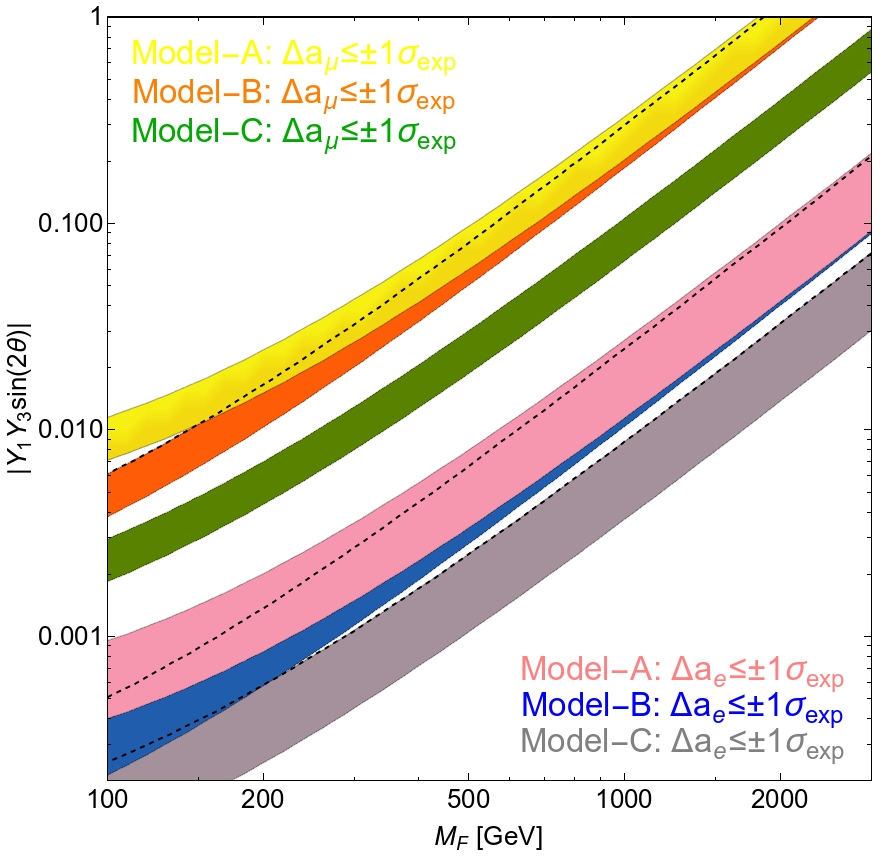}
\caption{Demonstration of the required Yukawa coupling to reproduce the experimentally observed lepton AMMs within $1\sigma$ values are presented as a function of vector-like fermion mass $M_F$. For the masses of two of the scalars running in the loop, we take $M_{F}+100$ GeV and $M_{F}+150$ GeV, respectively.  Here, for illustration, we have only plotted the dominant chiraly enhanced contribution; however, for numerical analysis, we have consider the full expressions. } \label{gM2}
\end{figure}
\textbf{Lepton magnetic dipole moment:}-- 
The NP contributions to lepton AMMs in this theory appear at the one-loop order, as shown in Fig.~\ref{fig:02} (Feynman diagram on the bottom). These contributions are typically large due to vector-like fermion mass insertion in the loop.  The outgoing photon in this Feynman diagram is emitted either by an internal scalar or fermion, or by both scalar and fermion, depending on the model. In Model-A, -B, and -C, scalars (fermions) running in the loop are  singly-charged (neutral), neutral (singly-charged), and doubly-charged (singly-charged)  states, respectively.  It is straightforward to evaluate the contribution arising from BSM states, which yields, 
\begin{align}
\Delta a_\ell&= -\frac{m_\ell}{4\pi^2} \bigg\{
Re\left( Y^{L\ast}_{kl}Y^R_{kl} \right) \frac{M_k}{M^2_{S^x_b}}G[r_{kb}]
\nonumber\\&+
\left( |Y^L_{kl}|^2+|Y^R_{kl}|^2 \right) \frac{m_\ell}{M^2_{S^x_b}} \widetilde G[r_{kb}]
\bigg\},\label{amm}
\end{align}
where summation over the BSM scalars and fermions must be understood. Furthermore, we have defined,
\begin{align}
&r_{kb}=\frac{M^2_k}{M^2_{S^x_b}},  
\\
&G[r]=f[r]+Q_\psi g[r],
\\
&\widetilde G[r]=\widetilde f[r]+Q_\psi \widetilde g[r],
\end{align}
with $Q_\psi=0,-1,+1$ for Model-A, -B, -C, respectively, and,
\begin{align}
&f[r]=2\widetilde g[r]=\frac{r^2-1-2r\ln r}{4(r-1)^3}, 
\\
&g[r]=\frac{r-1-\ln r}{2(r-1)^2},
\\
&\widetilde f[r]=\frac{2r^3+3r^2-6r+1-6r^2\ln r}{24(r-1)^4}.
\end{align}
Finally, the re-defined Yukawa couplings $(Y^\ast_L, Y_R)$ appearing in Eq.~\eqref{amm} are given by,
\begin{align}
&\textrm{Model-A}:  \nonumber \\& 
\eta_1: (Y_1 c_{\theta_+}, Y_3 s_{\theta_+}),\;\;
\eta_2: (-Y_1 s_{\theta_+}, Y_3 c_{\theta_+}), 
\\
&\textrm{Model-B}:  \nonumber \\& 
S_1: (Y_1 c_{\theta_0}, Y_3 s_{\theta_0}),\;\;
S_2: (-Y_1 s_{\theta_0}, Y_3 c_{\theta_0}), 
\\
&\textrm{Model-C}:  \nonumber \\& 
\kappa_1: (Y_1 c_{\theta_{++}}, Y_3 s_{\theta_{++}}),\;\;
\kappa_2: (-Y_1 s_{\theta_{++}}, Y_3 c_{\theta_{++}}). 
\end{align}

In Fig.~\ref{gM2}, we present beyond SM contributions to the muon and the electron anomalous magnetic moments arising in three versions of models under consideration. It is clear from this plot that the required large contributions as observed in the experiments can be naturally provided within this framework without requiring large Yukawa couplings. Furthermore, opposite signs of the muon and the electron AMMs can be incorporated via an appropriate choice of the signs of the associated Yukawa couplings.

\section{Experimental constraints}\label{sec:EXP}
This section briefly describes the phenomenological implications of the proposed models and the current experimental bounds on the BSM states, along with future collider prospects.  

\textbf{LHC bounds on scalars and fermions:}-- 
Model-A contains a DM candidate (see Sec.~\ref{sec:DM} for details) via which it can be tested in colliders. Specifically, the singly charged scalars can be efficiently pair-produced at the LHC through the $s$-channel $\gamma/Z$ exchange. Once pair-produced, each of them decays into a DM and a SM lepton, i.e., $pp\to \ell^+\ell^-+\slashed{E}_T$. This process mimics the standard  slepton searches carried out by ATLAS as well as CMS collaborations~\cite{Aad:2014yka,Sirunyan:2018nwe,Sirunyan:2018vig} and   non-observation of any such processes lead to a bound of $m_{S^\pm_i} \geq 450$ GeV~\cite{Sirunyan:2018nwe}.  

On the contrary, Model-B/C does not contain a DM candidate. Consequently, collider probes of these models are distinct from Model-A. Following Model-A, we assume that BSM scalars are heavier than BSM fermionic states in Model-B/C. Then pair-produced singly  (singly and doubly) charged scalars in Model-B (Model-C) decay into a pair of SM lepton ($\nu\nu$ or $\ell^+\ell^-$ depending on singly or doubly charged scalar) and a pair of BSM singly charged fermions ($\psi^+\psi^-$). In fact, the singly charged fermions also get pair-produced through the $s$-channel $\gamma/Z$ exchange, which provides the relevant bounds for these models. Note, however, that for a general charge assignment with an arbitrary value of $\beta$, a renormalizable coupling  responsible for the decay of these fermions may not be present; hence $\psi^\pm$ are expected to be long-lived.

To make them decay, we fix the $B-L$ charge such that $\beta=-1$ for Model-B/C, therefore a mixing term of the form $\mathcal{L}\supset m^\prime \overline \psi_L \ell_R$ is allowed. Its contribution to SM lepton masses can be fully neglected if $\epsilon\ll 1$, where we define $m^\prime\equiv \epsilon v_H/\sqrt{2}$. Through this mixing, the vector-like leptons will promptly decay if $\epsilon \gtrsim 2\times 10^{-7}$~\cite{Bhattiprolu:2019vdu}. For a quasi-stable vector-like lepton, assuming chargino like interactions, ATLAS search~\cite{ATLAS:2019gqq} provides a bound of $m_{\psi^\pm}\geq 750$ GeV~\cite{Bhattiprolu:2019vdu}. On the other hand, for the prompt decay scenario, each of the pair-produced vector-like lepton decays into $\psi\to h\ell$, $\psi\to Z\ell$, and $\psi\to W\nu_\ell$, for which currently there is no collider bound. However, HL-LHC (14 TeV with integrated luminosity of $3 ab^{-1}$) will probe these processes and should be able to exclude up to about $m_{\psi^\pm}\geq 460$ GeV for the first two generations~\cite{OsmanAcar:2021plv} and  $m_{\psi^\pm}\geq 600$ GeV for tau-like $\psi$~\cite{Bhattiprolu:2019vdu}.          

\textbf{Electroweak precision constraints:}-- 
Neutrino mass and BSM contributions to the lepton anomalous magnetic moments heavily depend on the splitting between the neutral (and charged) BSM scalars; hence electroweak precision measurements provide stringent constraints on the model parameters. The so-called $T$-parameter is the most crucial among these oblique parameters, which we  compute following Refs.~\cite{Peskin:1990zt,Barbieri:2006dq, Grimus:2008nb,Funk:2011ad}  and allow it to vary within the range given by~\cite{ParticleDataGroup:2020ssz},    
\begin{align}
\Delta T=0.03\pm 0.12.    
\end{align}

\textbf{LEP bounds on vector boson:}--  
Since we study gauged $B-L$ extension of the SM, this theory contains a heavy gauge boson $Z^\prime$, which is significantly constrained from the non-observation of any direct signature at LEP and LHC. This gauge boson couples to quarks as well as leptons, thus $Z^\prime$ can be produced and searched for at the LEP via $s$-channel $e\overline e\to Z^\prime \to f\overline f$  processes. To assure LEP-II bound~\cite{LEP:2003aa}, we impose 
\begin{align}
\frac{M_{Z^\prime}}{g^\prime} > 6.94 \textrm{TeV at $95\%$ C.L.},    
\end{align} 
which is derived from effective four-lepton operator~\cite{Carena:2004xs} and valid for $M_{Z^\prime}\gg 200$ GeV. 

\textbf{LHC bounds on vector boson:}-- 
Furthermore, at hadron colliders, $Z^\prime$ can be efficiently produced in the $s$-channel due to its couplings to quarks, which would show up as resonances in the invariant mass distribution of the decay products.   Searching for a massive resonance at the LHC decaying into di-lepton/di-jet imposes stringent limits on the respective production cross-section. The current data from 13 TeV LHC search for a heavy resonance decaying into two leptons (assuming a $100\%$ branching ratio) via $pp\to Z^{\prime}\to \ell^+\ell^-$ provides the tightest constraint of $M_{Z^\prime} > 4.9$ TeV. This bound is somewhat relaxed for other branching ratios, which is depicted in Fig.~\ref{fig:DM03} using the current data (ATLAS~\cite{ATLAS:2019erb} and CMS~\cite{CMS:2021ctt}) and future projections (HL-LHC~\cite{Ruhr:2016xsg} and FCC-hh~\cite{Helsens:2019bfw}) in  the  coupling versus mass plane~\cite{Padhan:2022fak}.

\textbf{Cosmological constraints on vector boson:}-- 
Since neutrinos are Dirac particles in our scenario, the existence of right-handed neutrinos $\nu_R$ is implied. Since these right-handed neutrinos are mass degenerate with left-handed neutrinos $\nu_L$, they could contribute to the effective number of relativistic degrees of freedom $N_{eff}$ in the early Universe.  In the case of purely SM interactions, the contributions are completely negligible. However, in our model, $\nu_R$ have gauge interactions with $Z^\prime$, through which they can be in thermal equilibrium with the SM plasma in the early Universe via $s$-channel $f\overline f \leftrightarrow \nu_R\overline \nu_R$ processes that increases $N_{eff}$. Cosmological data, however, requires that   $\nu_R$s decouple from the SM plasma much earlier than the $\nu_L$. To be specific, Planck 2018 data~\cite{Planck:2018nkj,Planck:2018vyg} requires that $\nu_R$s must decouple at temperatures greater than $T> 0.6$ GeV~\cite{Abazajian:2019oqj}. 

The best current measurement implies $N_{eff}=2.99\pm 0.17$~\cite{Planck:2018nkj,Planck:2018vyg}, which is in complete agreement with the SM prediction $N^{SM}_{eff}=3.045$~\cite{Mangano:2005cc,Grohs:2015tfy,deSalas:2016ztq}. Then a $2\sigma$ C.L. corresponds to $\Delta N_{eff}< 0.285$, which we adopt in our analysis. Future sensitives of CMB-S4 experiments will reach a precision of $\Delta N_{eff}\sim 0.03$~\cite{CMB-S4:2016ple,Abazajian:2019eic} that can probe NP scale up to about 50 TeV~\cite{Abazajian:2019oqj,Luo:2020sho}.  For heavy $Z^\prime$, i.e., $M_{Z^\prime}> 20$ GeV, assuming that $\nu_R$ decouples before $T\sim 0.6$ GeV, one obtains $M_{Z^\prime}/g^{\prime}> 11.4$ TeV, for the case of the standard $B-L$ theory~\cite{Luo:2020sho} (this limit becomes much stronger in the lower mass range, see, e.g., Ref.~\cite{Abazajian:2019oqj}). We re-scale this bound using the unconventional charge assignment of the right-handed neutrinos in our theory, and the corresponding Planck 2018 bound is presented in Fig.~\ref{fig:DM04}.

\begin{figure}[b!]
\includegraphics[width=0.3\textwidth]{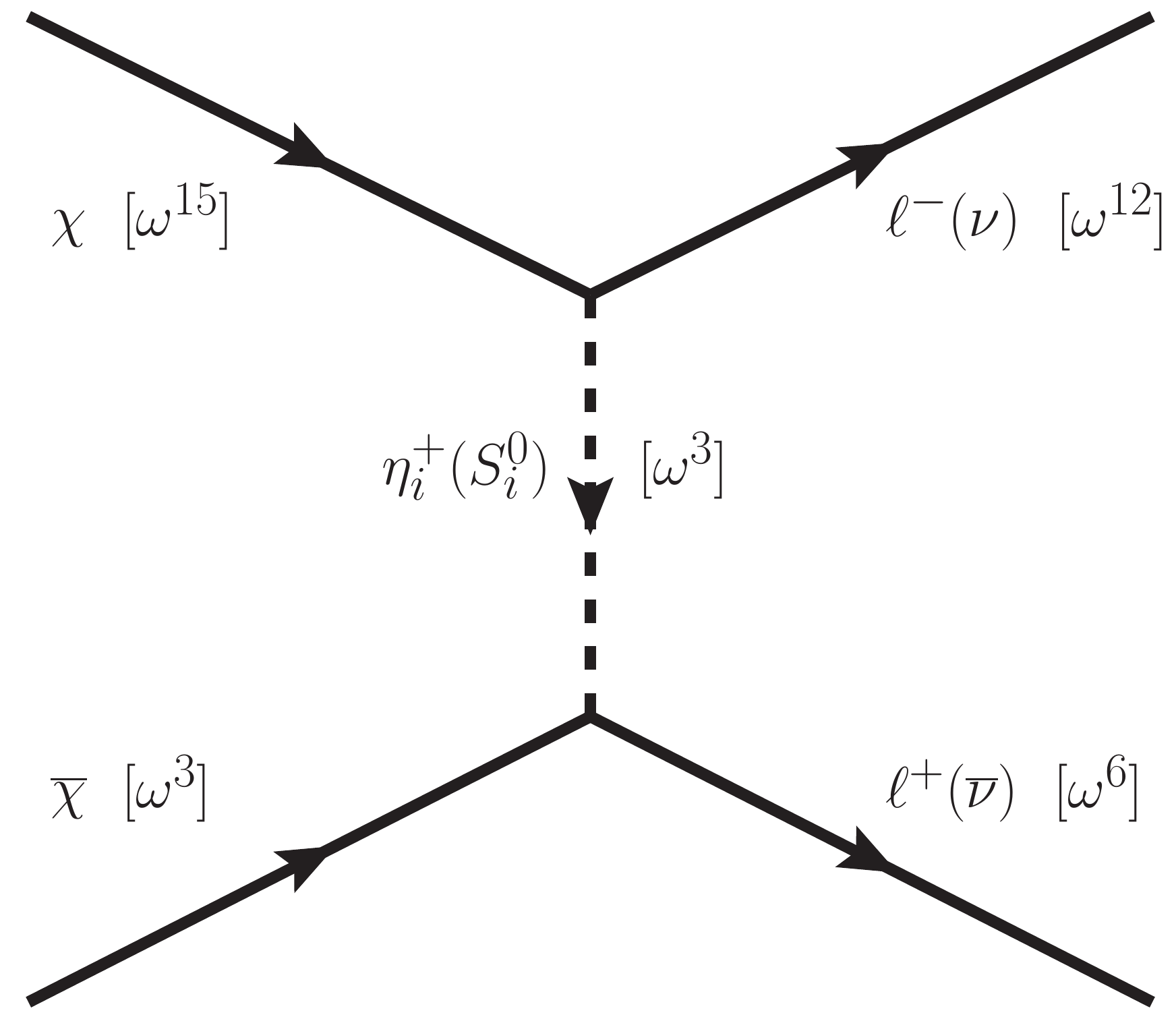}
\caption{Dark matter annihilation into SM leptons.} \label{fig:annihilation}
\end{figure}
\section{Dark Matter}\label{sec:DM}
In this theory, the Diracness of neutrinos is protected by the $U(1)_{B-L}$ symmetry. Remarkably, this same symmetry is also responsible for stabilizing the DM candidate. The spontaneous symmetry breaking of $U(1)_{B-L}$ by the VEV of $\sigma$ leaves a residual discrete symmetry $\mathcal{Z}_D$ (such that $\omega^D=1$) and stabilizes the DM. In particular, stability is guaranteed if $D>2$ as well as if $D$ is even \cite{Bonilla:2018ynb}.    In such a case, all SM fields transform as even powers of $\omega$, and the lightest particle transforming as an odd power of $\omega$ will be automatically stable. For concreteness, for Model-A, we fix $\beta=-1/2$ in our analysis, and the corresponding charges of all fields under the residual symmetry are presented in Table~\ref{tab:Dark}.  As can be seen from this Table, all particles in the dark sector ($\psi, \phi, S^0, \eta^+$) carry charges that are odd powers of $\omega$. 

\begin{table}[t!]
\centering
{\footnotesize
\resizebox{0.35\textwidth}{!}{
\begin{tabular}{|c|c|c|}
\hline
Fields &  $U(1)_{B-L}$  &  $U(1)_{B-L}\to \mathcal{Z}_{18}$  \\[3pt]
\hline\hline
$Q_L, u_R, d_R$&
$\frac{1}{3}$ & $\omega^2$ \\[3pt]
\hline
$L_L, \ell_R$&
$-1$ & $\omega^{12}$ \\[3pt]
\hline
$\nu_R$&
$-4,5$ & $\omega^{12}$ \\[3pt]
\hline
$\psi$&
$-\frac{1}{2}$ & $\omega^{15}$ \\[3pt]
\hline \hline

$H$&
$0$ & $\omega^{0}$      
 \\[3pt]
\hline
$\sigma$&
$3$ & $\omega^{0}$      
 \\[3pt]
\hline
$\phi^\ast$&
$\frac{1}{2}$ & $\omega^{3}$      
 \\[3pt]
\hline
$S^0$&
$\frac{7}{2}$&$\omega^{3}$ 
 \\[3pt]
\hline
$\eta^+$&
$\frac{1}{2}$&$\omega^{3}$ 
 \\[3pt]
\hline 
\end{tabular}
}
    \caption{Charges of particles under the residual symmetry in Model-A with $\beta=-1/2$. The dark sector transforms as an odd power of $\omega$, whereas the rest of the particles transform as even powers of $\omega$.  See text for details. }
    \label{tab:Dark}
    }
\end{table}

The physical states in the dark sector of the Model-A consists of three Dirac fermions: $\psi_{i},\,(i=1-3)$, two singly-charged scalars, $\eta_{i}$ and two complex neutral scalars, $S_{i}$ with $i=1,2$. The DM candidate is the lightest Dirac fermion, which we choose  to be $\psi_{2}\equiv \chi$. A typical DM annihilation channel in our model is demonstrated in Fig.~\ref{fig:annihilation}. Before delving into the thermal freeze-out scenario of the DM, we delineate the relevant parameter space and constraints set by the DM direct detection experiments.

\subsection{DM Direct Detection and Parameter Space}
\textbf{DM Direct Detection:}--
The fermionic DM, being charged under $U(1)_{B-L}$, will scatter with the nucleon via the exchange of $Z'$ at the tree-level, and in the limit of zero momentum transfer one can deduce the following effective interaction term in the Lagrangian between the DM and the nucleon,
\begin{equation}
    -{\cal L}\supset\frac{g'^{2}}{2 M_{Z'}^{2}}\overline\chi\gamma_{\mu}\chi \overline{N}\gamma^{\mu}N\;,
    \label{DM-N}
\end{equation}
where, $N$ is the nucleon. The spin independent scattering cross-section associated with this effective interaction is given as,
\begin{equation}
    \sigma_{\mathrm{SI}}=\frac{1}{4\pi}\frac{g'^{4}\mu_{r}^{2}}{4 M_{Z'}^{4}}\;,
    \label{SIcross}
\end{equation}
where, the reduced mass is $\mu_{r}=\frac{m_{N}M_{\mathrm{DM}}}{m_{N}+M_{\mathrm{DM}}}$, and the cross-section is insensitive to the DM mass. Consequently, for a fixed DM mass, the limits on the $\sigma_{\mathrm{SI}}$ set by the currently operating DM direct detection experiments will constrain the region of two dimensional $M_{Z'}-g'$ parameter plane as shown in Fig.~\ref{fig:DM03}.

\textbf{DM Parameter Space:}--
We want to correlate the constraints on the electron and muon magnetic dipole moments with the parameter space associated with the DM relic density. Hence, based on the choice of the Yukawa matrices given in Eq.~(\ref{Ys}), and the scalar masses and mixing angles given in Eq.~(\ref{scalarmassmix}), we choose the parameter space for the DM relic density analysis which is spanned by 
$M_{\mathrm{DM}}$, $M_{F_{1}}$, $M_{\eta_{1}}$, $M_{\eta_{2}}$, $M_{S_{1}}$, $M_{S_{2}}$, $\theta_{+}$, $\theta_{0}$,
$Y_{1_{ee}}$, $Y_{1_{\mu\mu}}$, $Y_{3_{ee}}$, $Y_{3_{\mu\mu}}$, $g'$ and $M_{Z'}$. Besides, the Yukawa couplings $Y_{2}$ being connected with the neutrino mass generation are much smaller than $Y_{1}$ and $Y_{3}$, and therefore don't play any significant role in the DM relic density analysis.

\subsection{DM Relic Density}
We consider the standard thermal freeze-out to achieve the correct DM relic density, $\Omega h^{2}=0.12\pm0.001$ ($68\%$ C.L.) observed by Planck \cite{Planck:2018vyg}. The dominant (co)annihilation processes which control the DM relic density for our considered DM parameter space, where $Y_{1}$ and $Y_{3}$ Yukawa matrices being diagonal, are enumerated in the following.

\begin{itemize}
    \item $\overline\chi\,\chi\rightarrow \mu^{+}\mu^{-}\,\,\mathrm{and}\,\,\overline{\nu}_{\mu}\nu_{\mu}$ via the exchange of $\eta^{+}_{1,2}$ and $S_{1,2}$ at the t-channel, respectively.
    
    \item $\overline\chi\,\chi\rightarrow \overline{q}q,\,l^{+}l^{-},\,\overline{\nu}_{l}\nu_{l}$ i.e. to the SM quark ($q$), charged lepton ($l$) and neutrino ($\nu_{l}$) pairs via the exchange of $Z'$ at the s-channel.
\end{itemize}

When the mass splitting between the DM candidate and another dark sector particle carrying same dark charge is small, the coannihilation channels also become important. In our case, the dominant coannihilation channels involving the DM candidate $\chi$ and other dark sector particles: $\psi_{1}$, $\eta^{+}_{1,2}$ and $S_{1,2}$ are,
\begin{itemize}
    \item $\chi\,\overline{\psi}_{1}\rightarrow \mu^{-}e^{+},\,\nu_{\mu}\overline{\nu}_{e}$, $\chi\,\eta^{-}_{1,2},\,\chi\,S^*_{1,2}\rightarrow \overline{X'}X$ where $X,\,X'$ are the SM particles, and the conjugated channels.
    \item $\psi_{1}\,\eta^{-}_{1,2},\,\psi_{1}S^{*}_{1,2},\,\eta^{+}_{i}\,\eta^{-}_{j},\, S_{i}\,S^{*}_{j},\,\eta^{+}_{i}\,S^{*}_{j}\rightarrow\overline{X'}X$ where $i,\,j=1,\,2$, and the relevant conjugated channels.
\end{itemize}

As the DM relic density involves multiple (co)annihilation channels, which also depend on the multi-dimensional parameter space, we consider two scenarios to capture the dynamics in a simplified manner. First, we consider the mass of the $Z'$ quite heavy compared to DM and other BSM particles so that the contribution of the $Z'$ mediated channels in the thermal freeze-out become negligible. We call it the {\it $Z'$ decoupled scenario}, and in this case, the thermal freeze-out of the DM is controlled by the BSM Yukawa sector of the Model-A. In the second scenario, dubbed here as {\it BSM Yukawa + $Z'$ gauge boson scenario},  we take into account both the simultaneous contributions of the BSM Yukawa sector and the $Z'$ gauge bosons to determine the DM relic density.

\begin{figure}[th!]
    \centering
    \includegraphics[width=0.45\textwidth]{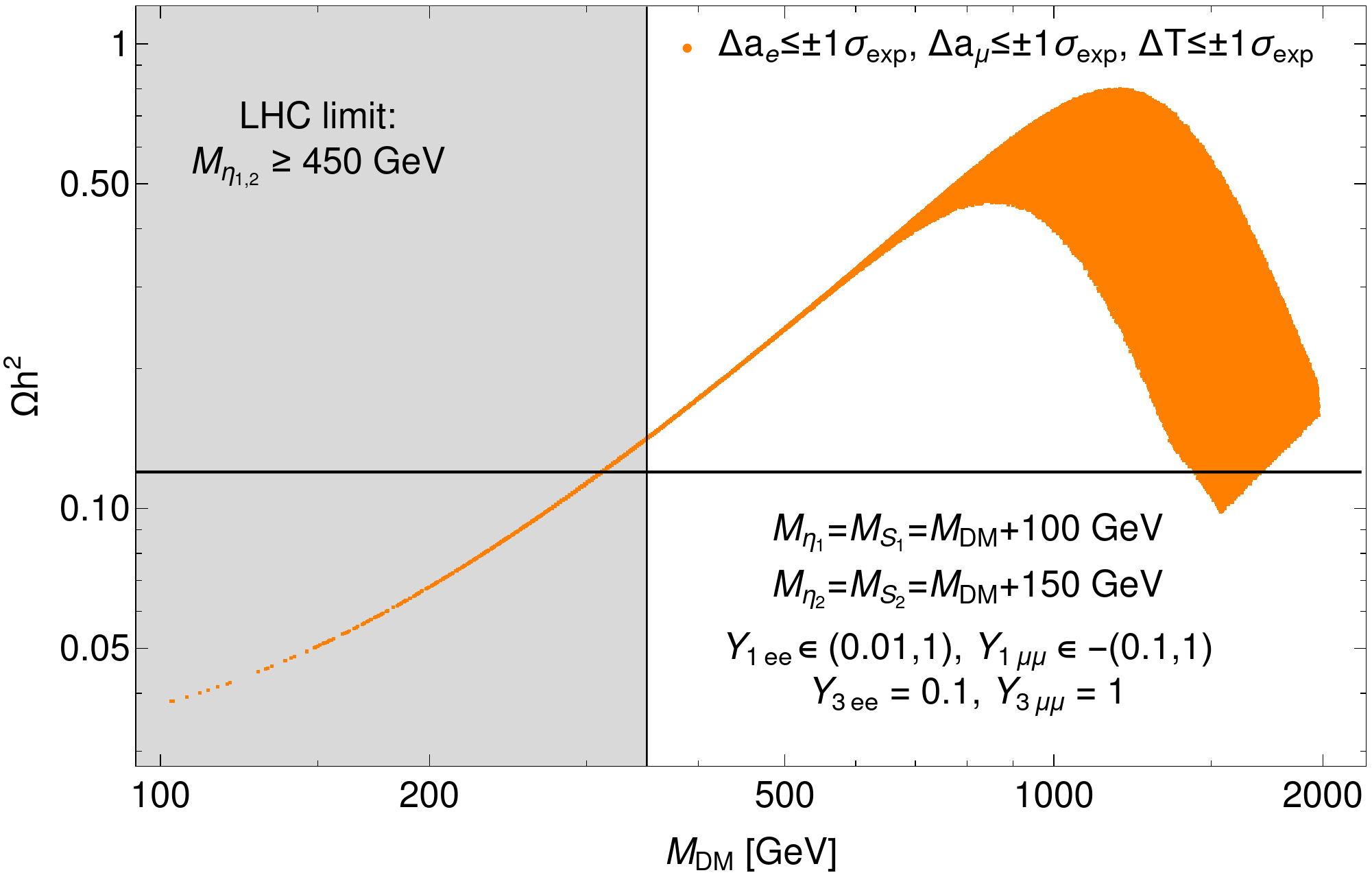}
    \caption{The correlation between the mass of the DM candidate $F_{2}$, $M_{DM}$ and the DM relic density $\Omega h^{2}$ for the parameter space which satisfy the constraints on the electron and muon magnetic moments, $\Delta a_{e}$ and $\Delta a_{\mu}$ and T-parameter $\Delta T$ for the $Z'$ decoupled scenario. The black horizontal line represents the correct DM relic density determined by Planck.}
    \label{fig:DM01}
\end{figure}

\textbf{$Z'$ decoupled scenario:}--
Although we consider the $Z'$ contribution to be negligible in determining the DM relic density for this scenario as mentioned above, the relevant parameter space is still large enough to disentangle the effects of the BSM Yukawa couplings, $Y_1$ and $Y_{3}$, the mass-splittings between the DM and the charged scalars $\eta_{1,2}^{+}$ and neutral scalars $S_{1,2}$, and the scalar mixing angles $\theta_{+}$ and $\theta_{0}$ on the DM relic density concretely just by scanning over the parameters randomly. Therefore, we consider a few benchmark points of the parameter space and discuss the impact of the variations of the parameters on the DM relic density.

First we consider a benchmark point where we fixed the following parameters as follows,
\begin{align}
    &M_{F_{1}}=M_{DM}+1\,\mathrm{GeV},\,M_{\eta_{1}}=M_{S_{1}}=M_{DM}+100\,\mathrm{GeV},\nonumber\\
    &M_{\eta_{2}}=M_{S_{2}}=M_{DM}+150\,\mathrm{GeV},\, \theta_{+}=\theta_{0}=0.7,\nonumber\\
    &Y_{3_{ee}}=0.1,\,Y_{3_{\mu\mu}}=1,
    \label{param}
\end{align}
and vary only the Yukawa couplings, $Y_{1_{ee}}\in (0.01,1)$ and $Y_{1_{\mu\mu}}\in-(0.1,1)$ randomly to select the points which simultaneously satisfy the constraints on the electron and muon magnetic moments $\Delta a_{e}$ and $\Delta a_{\mu}$ and T-parameter $\Delta T$. Afterwards, we calculate the relic density  of the DM $F_{2}$ for each of these selected points using micrOMEGAsv$5.2$ \cite{Belanger:2020gnr} with the model files generated by FeynRules \cite{Alloul:2013bka}.

\begin{widetext}
\begin{figure*}[th!]
\centering
    \centerline{\includegraphics[width=0.45\textwidth]{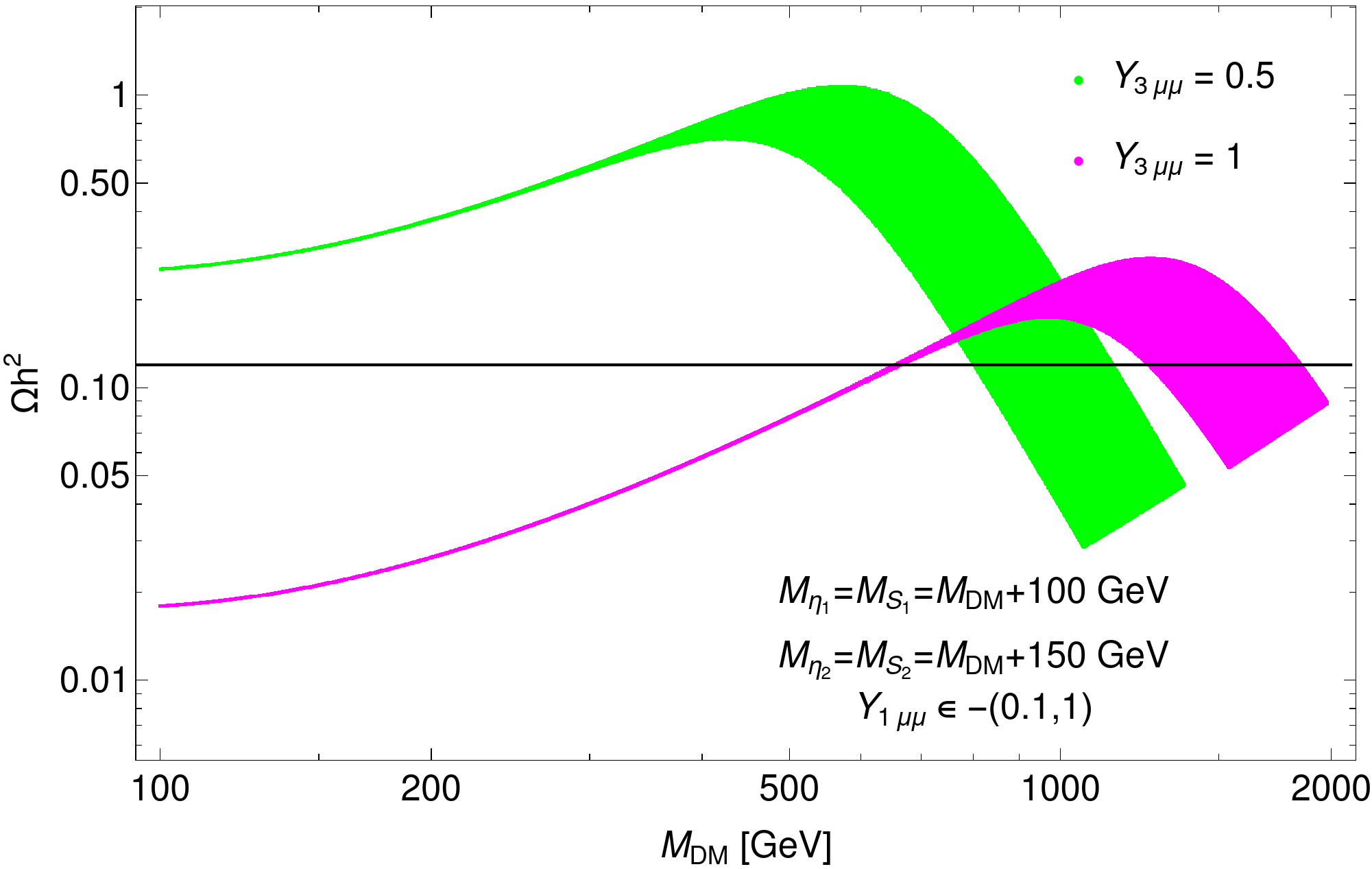}\hspace{2mm}\includegraphics[width=0.45\textwidth]{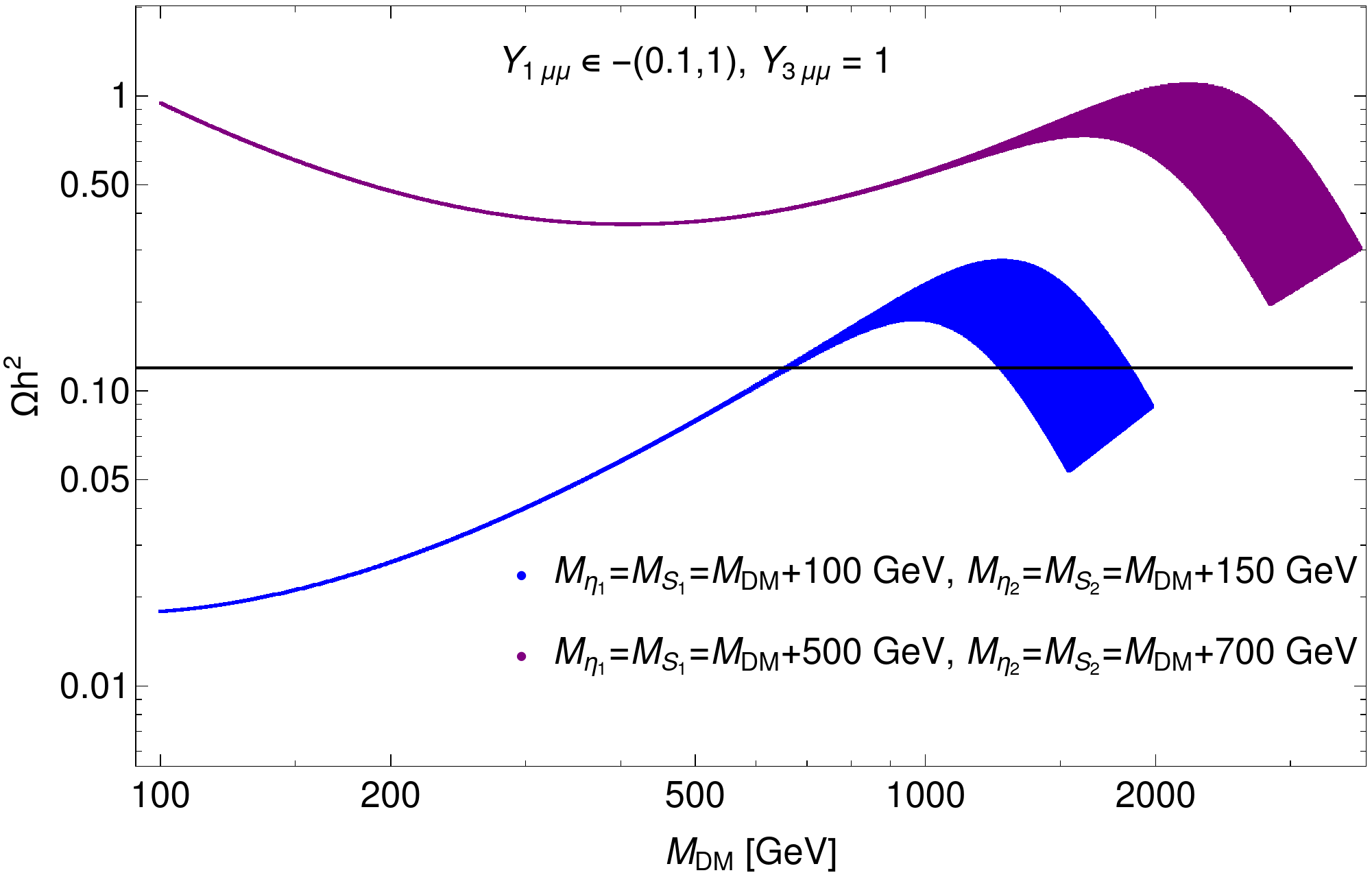}}
    \caption{The dependence of the DM relic density on the variation of $Y_{3_{\mu\mu}}$ for a fixed mass splitting (left fig.) and on the mass splittings between the DM and charged and neutral scalars for a fixed value of $Y_{3_{\mu\mu}}$ (right fig.). Here we vary $Y_{1_{\mu\mu}}\in -(0.1,1)$. For simplicity, in this presentation, we only consider the parameter points satisfying the constraints on the muon magnetic moment $\Delta a_{\mu}$ and the T parameter $\Delta T$ simultaneously. Again the black horizontal line represents the correct DM relic density. The respective collider bounds are not shown in this plots, see text for details.}
    \label{fig:DM02}
\end{figure*}
\end{widetext}

Subsequently, for the parameter set given in Eq.~\eqref{param}, the correct DM relic density is obtained for $M_{DM}=310$ GeV which is excluded by the LHC limits, and for $1.4-1.7$ TeV mass range as seen in Fig.~\ref{fig:DM01}. To illustrate it, first, we write down the representative annihilation cross-section at low-velocity approximation that is associated with the DM, $\chi$ annihilating into the SM leptons (charged and neutrinos) via the exchange of charged ($\eta_{1,2}$) and neutral ($S_{1,2}$) scalars at t-channel,
\begin{equation}
    \sigma v_{0}=\frac{M_{DM}^2\left(\tilde{Y}^{2}_{1_{a}}+\tilde{Y}^{2}_{3_{a}}\right)^{2}}{32\pi \left(M_{DM}^{2}+M_{\tilde{S}_{i}}^{2}\right)^2}\;,
    \label{anncross}
\end{equation}
where the terms having small masses of the final-state leptons are neglected, and $\tilde{Y}_{1_{a}}$ and $\tilde{Y}_{3_{a}}$ are the Yukawa couplings of electron or muon sector (here denoted by $a$) redefined by absorbing the appropriate scalar mixing angles associated with charged or neutral scalar (here expressed with $\tilde{S}_{i}$). As we choose $Y_{3_{\mu\mu}}=1$ for our parameter set in Eq.~\eqref{param}, when the DM mass is in $100-680$ GeV range, the term like $\tilde{Y}_{3_{\mu\mu}}^{4}M_{DM}^{2}$ in the numerator of Eq.~\eqref{anncross} dominates the cross-section, and therefore we see the narrow band for that mass range. In contrast, when $M_{DM}>680$ GeV, the DM mass is large enough to make the terms like $\tilde{Y}_{1_{\mu\mu}}^{4}M_{DM}^{2}$ in the numerator of Eq.~\ref{anncross} also comparable to the terms with $Y_{3_{\mu\mu}}$, and thus we see the band at larger masses in the DM relic density plot as $|Y_{1_{\mu\mu}}|$ takes value in the range $(0.1,1)$. Besides, as the mass-splittings are small ($150$ GeV) when the DM mass is close to $O(\mathrm{TeV})$, the coannihilation channels also contribute significantly to the DM relic density, so the simplified Eq.~\ref{anncross} is not applicable for higher DM mass ranges with small mass-splittings.

\begin{widetext}
\begin{figure*}[th!]
\centering
\centerline{\includegraphics[width=0.45\textwidth]{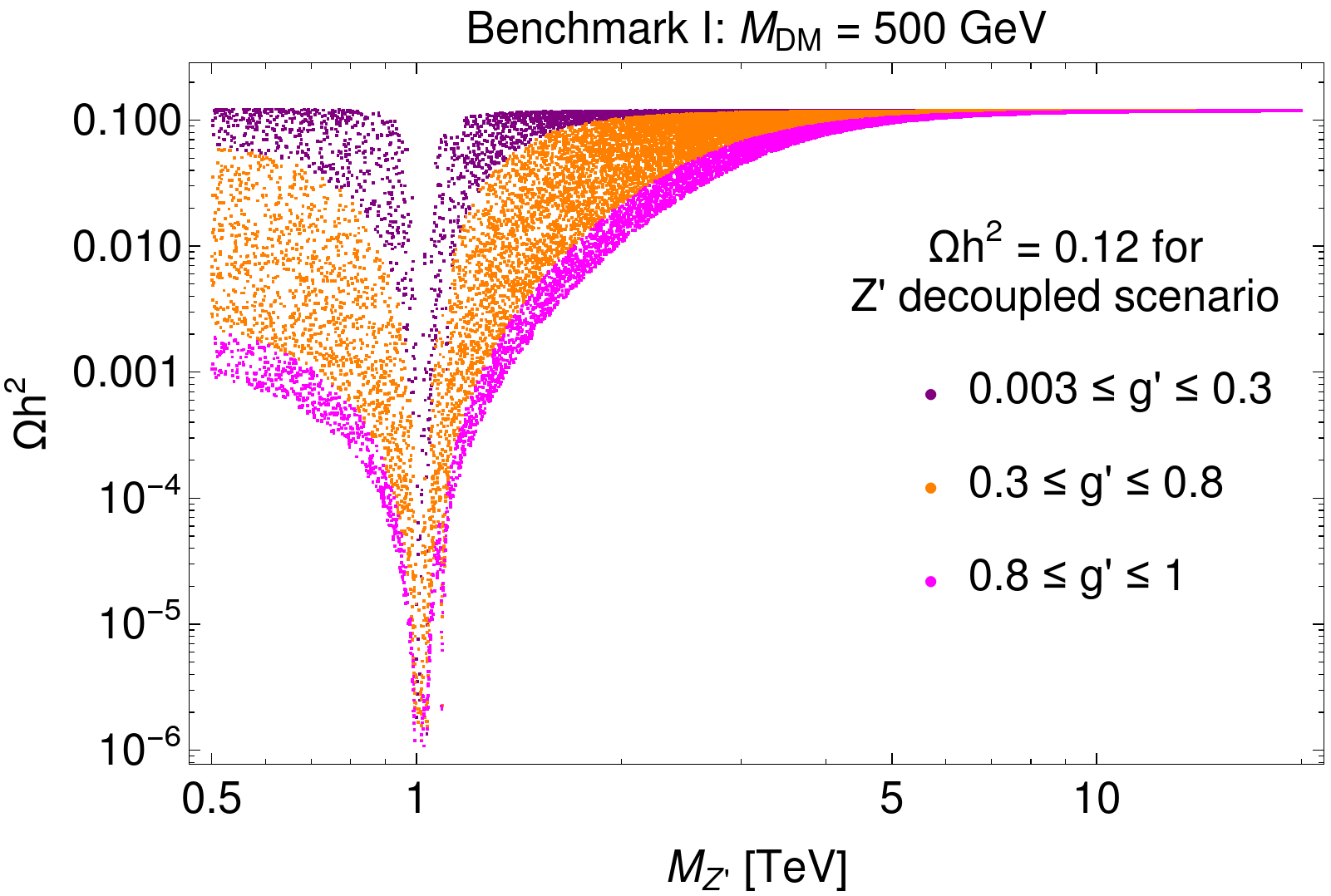}\hspace{0cm}\includegraphics[width=0.45\textwidth]{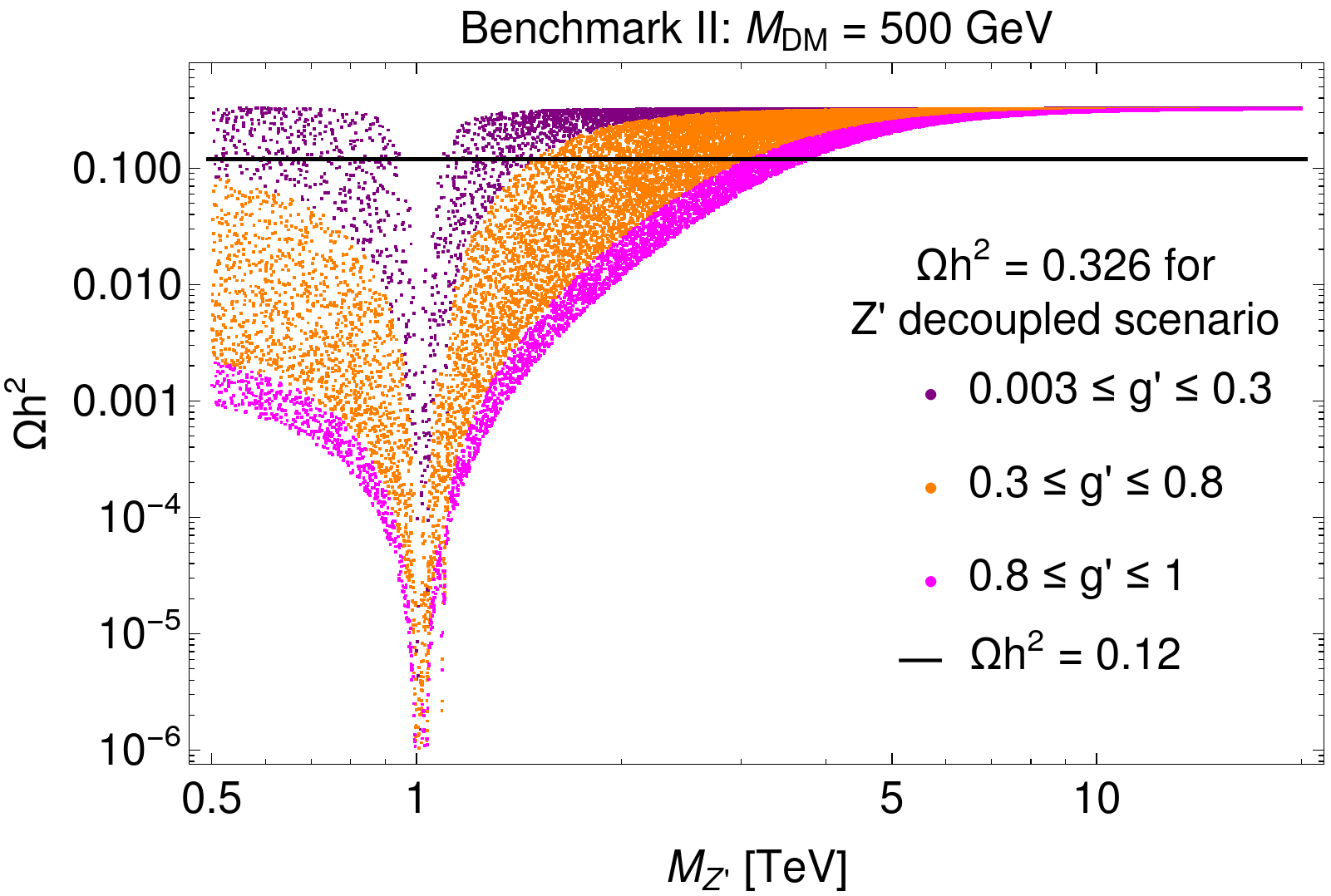}}
\vspace{1mm}
\centerline{\includegraphics[width=0.45\textwidth]{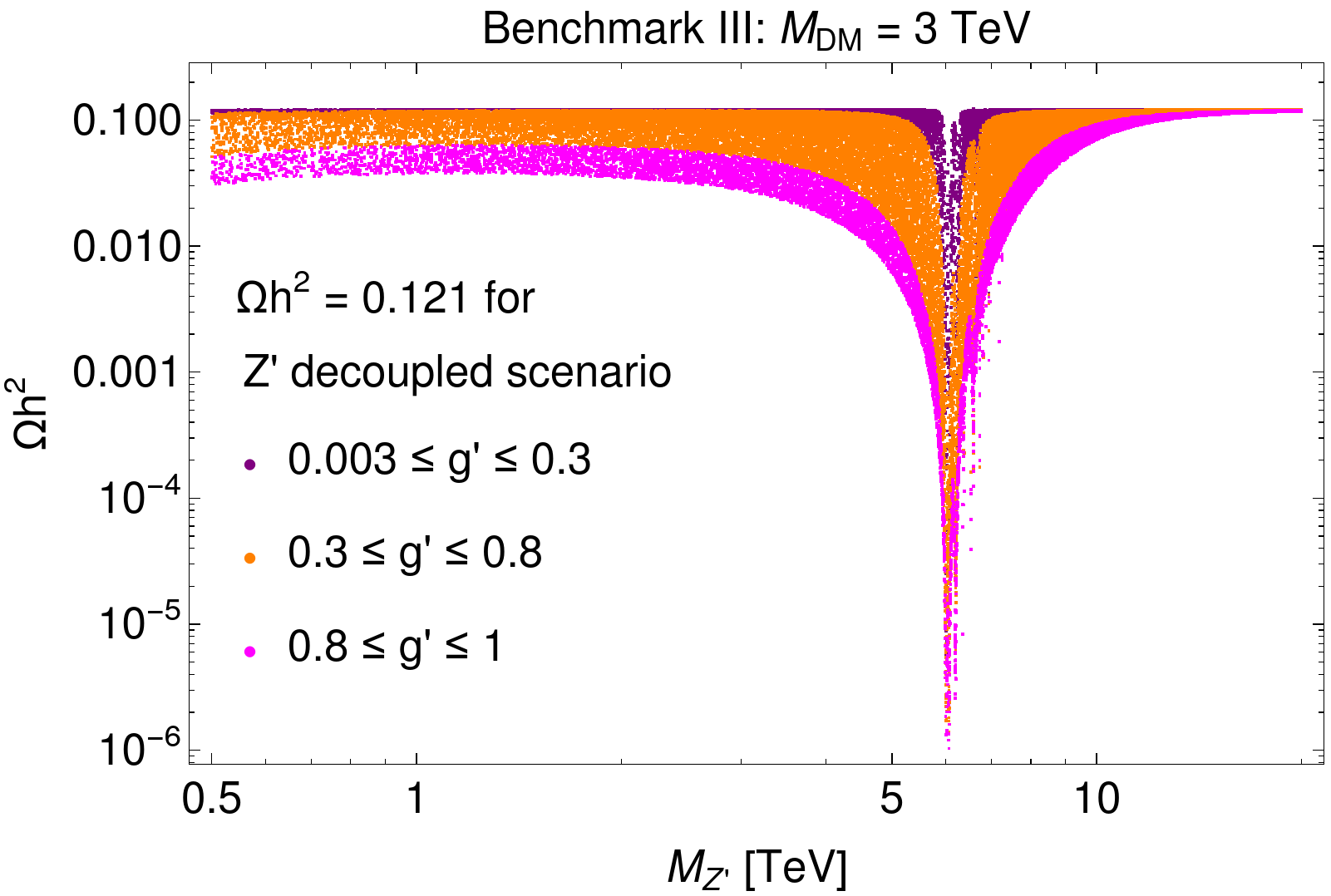}\hspace{0cm}\includegraphics[width=0.45\textwidth]{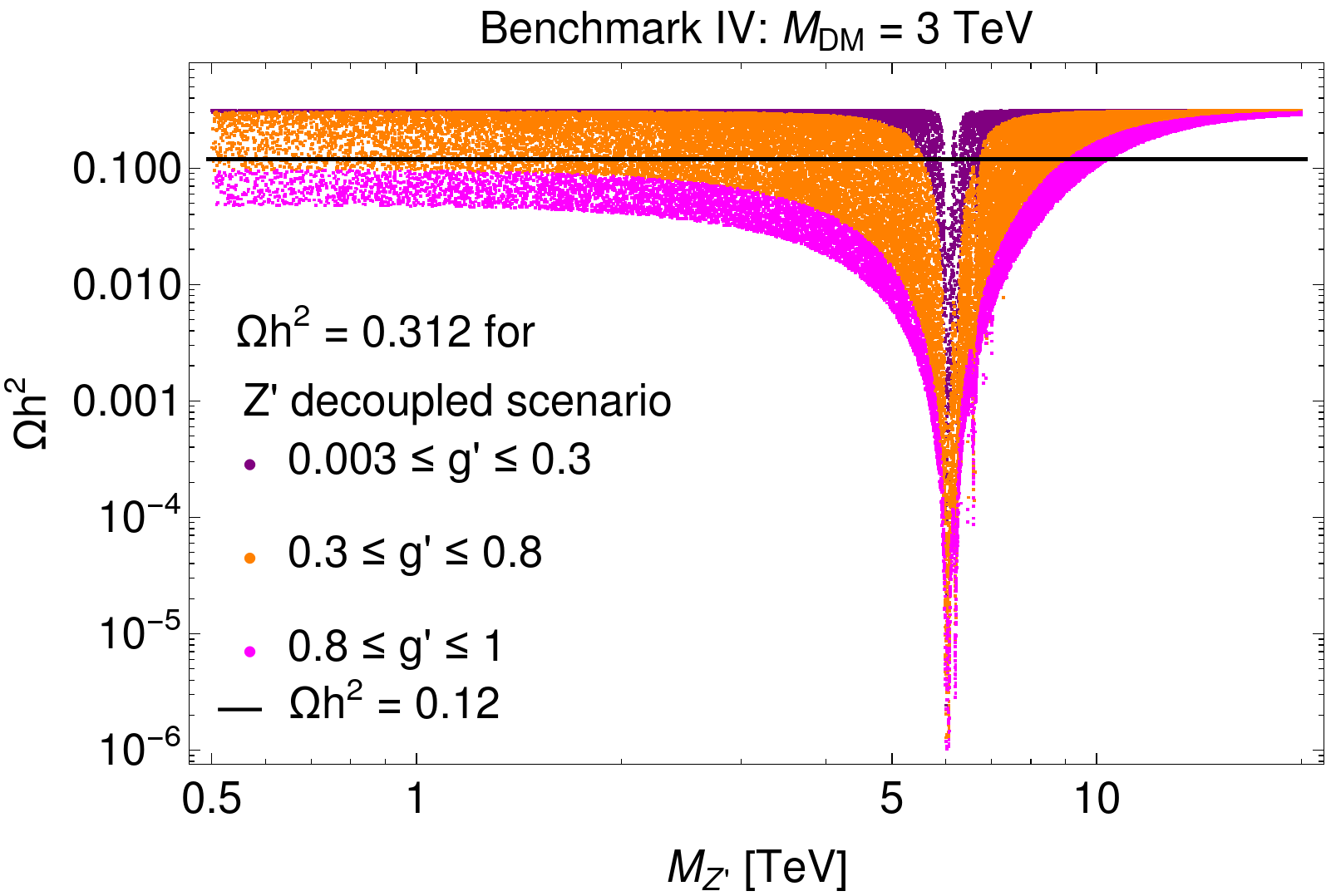}}
\caption{Correlation between the mass of the $Z'$ gauge boson $M_{Z'}$ and the DM relic density $\Omega h^{2}$ for four benchmark points given in Table \ref{benchmarktab}.}
\label{fig:DM03}
\end{figure*}
\end{widetext}

In Fig.~\ref{fig:DM02} (left), we can see that if $Y_{3_{\mu\mu}}$ is set to 0.5 instead of 1, the annihilation cross-section decreases for a wide range of the DM mass, and for this reason, the DM relic density remains overabundant for DM mass close to TeV scale as opposed to the case with $Y_{3_{\mu\mu}}=1$ for which it remains underabundant for similar DM mass range. In addition, when we increase the mass-splittings between the DM and the charged and neutral scalars, for lower mass range, the denominator in Eq.~\ref{anncross} becomes more significant, and the annihilation cross-section decreases, which results in the overabundant DM. Besides, larger mass-splittings suppress the coannihilation processes during the thermal freeze-out resulting in the overabundance of the DM for masses in the TeV range.

Furthermore, to capture the impact of $Z'$ on the DM relic density, first, we consider four benchmark points presented in Table. \ref{benchmarktab} which simultaneously satisfy the constraints on the electron and muon magnetic moments $\Delta a_{e}$ and $\Delta a_{\mu}$ and T-parameter $\Delta T$ in the $Z'$ decoupled scenario.
\begin{widetext}
\begin{table*}[t!]
    \centering
    \begin{tabular}{|c|c|c|c|c|c|c|c|c|c|c|c|c|c|c|}
    \hline
     Benchmark &  $M_{F_{2}}$ & $M_{F_{1}}$ & $M_{\eta_{1}}$ & $M_{\eta_{2}}$ & $M_{S_{1}}$ & $M_{S_{2}}$ & $\theta_{+}$ & $\theta_{0}$ & $Y_{1_{ee}}$ & $Y_{1_{\mu\mu}}$ & $Y_{3_{ee}}$ & $Y_{3_{\mu\mu}}$ & $\Omega h^{2}$\\
     \hline
     I & 500 GeV & 501 GeV & 961.46 GeV & 1357.04 GeV & 858.75 GeV & 1085.4 GeV & 0.05 & 0.56 & 0.09 & -1.33 & 0.67 & 0.36 & 0.12\\
     \hline
     II & 500 GeV & 501 GeV & 1008.3 GeV & 1983.3 GeV & 979.83 GeV & 1408 GeV & 0.04 & 0.02 & 0.09 & -1.07 & 0.3 & 0.54 & 0.326\\
     \hline
     III & 3000 GeV & 3001 GeV & 3160.5 GeV & 3582 GeV & 3077 GeV & 3672.2 GeV & 0.42 & 0.41 & 0.09 & -0.96 & 0.78 & 0.73 & 0.121\\
     \hline
     IV & 3000 GeV & 3001 GeV & 3124.6 GeV & 3562 GeV & 3205.6 GeV & 3611.4 GeV & 0.6 & 0.46 & 0.06 & -1.11 & 0.84 & 0.45 & 0.312\\
     \hline
    \end{tabular}
    \caption{Benchmark points in the $Z'$ decoupled scenario for $M_{DM} =500$ GeV and $M_{DM}=3000$ GeV.}
    \label{benchmarktab}
\end{table*}
\end{widetext}

For benchmark point I and II of Table \ref{benchmarktab} with $M_{DM}=500$ GeV, the thermal freeze-out of the DM is dominated by the processes, $\overline\chi\chi\rightarrow \overline{\nu}_{\mu}\nu_{\mu},\,\mu^{+}\mu^{-}$. On the other hand, for benchmark point III and IV with $M_{DM}=3000$ GeV, the thermal freeze-out processes are dominated by the coannihilation, for example, by $S_{1}S^{*}_{1}\rightarrow W^{+}W^{-},\,Z\,Z$ and $\eta^{+}_{1}\eta^{-}_{1}\rightarrow W^{+}W^{-},\,Z\,Z$, respectively.

\textbf{BSM Yukawa sector + $Z'$ gauge boson:}--
Now we calculate the DM relic density for the each benchmark points in Table \ref{benchmarktab} by varying the mass of the $Z'$ gauge boson $M_{Z'}\in (0.5,20)$ TeV and the gauge coupling $g'\in(0.02,1)$.
\begin{widetext}
\begin{figure*}[th!]
    \centering
    \includegraphics[width=1\textwidth]{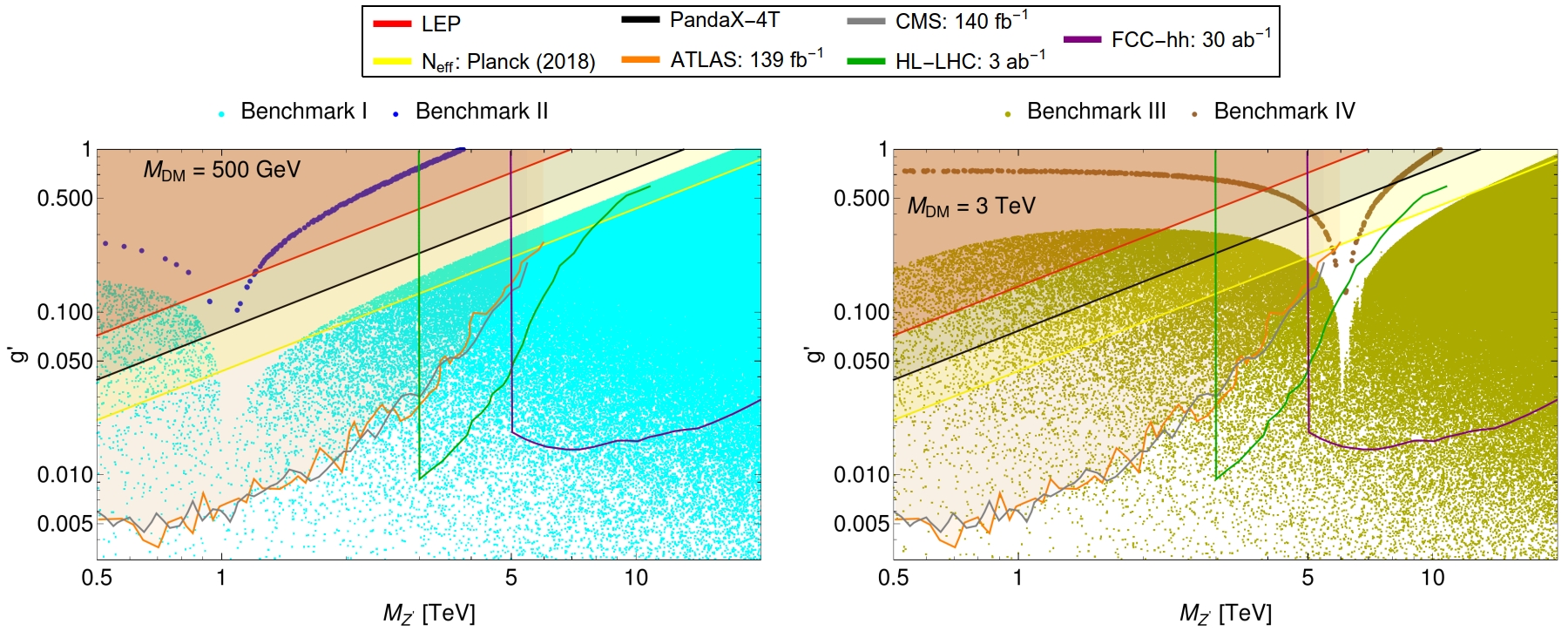}
    \caption{Correlation between $M_{Z'}$ and $g'$ for four benchmark points. For $m_{DM}=500$ GeV (left fig.), the cyan and blue points correspond to $(M_{Z'},\,g')$ pairs that satisfy the constraints on DM relic density for benchmark I and benchmark II points in the $Z'$ decoupled scenario, respectively, whereas for $m_{DM}=3$ TeV (right fig.), the yellow-green and brown points represent those with same attribute for  benchmark III and benchmark IV points, respectively. The shaded regions are ruled out by current collider bounds, cosmological constraints, and from dark matter direct detection. Future colliders will probe regions inside green and purple lines (not shaded).}
    \label{fig:DM04}
\end{figure*}
\end{widetext}

From Fig.~\ref{fig:DM03}, we can see that for a fixed $M_{Z'}$, the increase of the gauge coupling $g'$ decreases the DM relic abundance determined in the $Z'$ decoupled scenario. Therefore, even if the BSM Yukawa sector sets an overabundant DM relic density, one can achieve its observed value by adjusting the $Z'$ mass and coupling $g'$ as seen in Fig.~\ref{fig:DM03} (upper and lower right figures). Moreover, we can see that the lower the value of the gauge coupling $g'$, the lower the minimum value of the $M_{Z'}$ where the $Z'$'s contribution to the DM relic abundance becomes negligible. Besides, we see the dip in the DM relic abundance due to the resonant enhancement of DM annihilation processes involving the exchange of $Z'$ at the s-channel when $M_{Z'}\sim 2 M_{DM}$. 

In Fig. \ref{fig:DM04}, we present the correlation between $M_{Z'}$ and $g'$ for our four benchmark points with all relevant constraints set by the collider searches and DM direct detection experiments, the observed DM relic density, and cosmological limit on the extra radiation. By correlating the Figs. \ref{fig:DM03} (upper left and right) and Fig.~\ref{fig:DM04} (left), for $M_{DM}=500$ GeV, we see that the current limits provided by ATLAS and CMS have already ruled out the $(M_{Z'},\,g')$ values for which the $Z'$ gauge boson give significant contributions to the DM relic abundance.  Again the close inspection of Figs. \ref{fig:DM03} (lower left and right) and Fig.~\ref{fig:DM04} (right) for $m_{DM}=3$ TeV indicates that apart from the small parameter space close to the resonance point around $M_{Z'}\sim 6$ TeV, almost all the $(M_{Z'},\,g')$ values are ruled out by LEP, LHC, DM direct detection (e.g. PandaX-4T \cite{PandaX-4T:2021bab} which has set the most stringent limit on the spin independent DM-nucleon cross-section for 10 GeV - 10 TeV DM mass range), and Planck constraint on $\mathrm{N}_{\mathrm{eff}}$. Furthermore, rest of the parameter space where $Z^\prime$ contributions could play an important role in determining DM relic abundance will be fully  probed by the HL-LHC and future collider like FCC-hh.

Since the DM particle is inducing the anomalous magnetic dipole moment via mass-insertion, the chirality flip enhancement
is strongly correlated to the mass generation mechanism of the associated lepton and causes related loop contributions to its
mass. Hence, if $M_{DM}\gtrsim \mathcal{O}(3)$ TeV, typically, fine-turning of a large degree is required to adjust the lepton mass correctly. For details, see, e.g., \cite{Athron:2021iuf} and references therein. This is why in this work, we explored DM phenomenology in the range $100\,\mathrm{GeV}\simlt M_{DM}\simlt 3$ TeV.

\section{Conclusion}\label{sec:CONCLUSIONS}
The Fermilab's Muon $g−2$ experiment has recently confirmed the longstanding tension of the muon AMM. Furthermore, the recent precise measurement of the electron AMM at the Berkeley Lab shows deviations from the theoretical prediction. These two anomalies together strongly hint towards physics beyond the Standard Model. Besides, the origin of neutrino mass remains a mystery even after the groundbreaking discovery of neutrino oscillation about twenty-five years ago. Moreover, even though we know DM exists, we do not yet know what it is at a fundamental level.  

This work proposes a class of radiative Dirac neutrino mass models where neutrino mass arises at a one-loop level. Furthermore, NP states that participate in neutrino mass generation also run through the loops and significantly contribute to $(g-2)_{\mu, e}$. These large contributions arise due to chirality enhancements required to simultaneously explain the $(g-2)_{\mu}$ and $(g-2)_{e}$ data.    For completeness, we have studied three benchmark models, one of which (Model-A) offers a Dark Matter candidate  whose stability is naturally protected without imposing additional symmetries by hand.  For Model-A, we have performed a detailed numerical analysis to investigate the correlations among the common model parameters accommodating neutrino oscillation data, the muon, the electron g-2, and dark matter relic abundance. Parameters that generate non-zero neutrino mass also play a non-trivial role in explaining the muon and the electron $g-2$ simultaneously; furthermore, these same parameters take part in dark matter annihilations to the SM particles in reproducing the correct relic abundance. This model is subject to numerous constraints arising from colliders and cosmology, through which it can be probed in the current and upcoming experiments.  A detailed study of lepton flavor violation and electric dipole moments of the electron and the muon and their possible links to the puzzles resolved in this work is left for future work.  
\bibliographystyle{style}
\bibliography{reference}
\end{document}